\documentclass[twocolumn]{aastex62}

\received{October 9, 2018}
\revised{January 9, 2019}
\accepted{\today}
\submitjournal{ApJ}

\shorttitle{Metallicity Study in M83 with COS}
\shortauthors{Hernandez et al.}
\begin{document}

\title{THE FIRST METALLICITY STUDY OF M83 USING THE INTEGRATED UV LIGHT OF STAR CLUSTERS\footnote{Based on observations made with the Hubble Space Telescope under program ID 14681.}}

\correspondingauthor{Svea Hernandez}
\email{sveash@stsci.edu}

\author{Svea Hernandez}
\affiliation{Space Telescope Science Institute,
3700 San Martin Drive, 
Baltimore, MD 21218, USA}

\author{S{\o}ren Larsen}
\affiliation{Department of Astrophysics / IMAPP, Radboud University,
PO Box 9010, 
6500 GL Nijmegen, The Netherlands}

\author{Alessandra Aloisi}
\affiliation{Space Telescope Science Institute,
3700 San Martin Drive, 
Baltimore, MD 21218, USA}

\author{Danielle A. Berg}
\affiliation{Department of Astronomy, 
The Ohio State University, 
140 W. 18th Avenue, 
Columbus, OH 43202, USA}

\author{William P. Blair}
\affiliation{Center for Astrophysical Sciences, 
Department of Physics and Astronomy, 
The Johns Hopkins University, 
Baltimore, MD 21218, USA}

\author{Andrew J. Fox}
\affiliation{Space Telescope Science Institute,
3700 San Martin Drive, 
Baltimore, MD 21218, USA}

\author{Timothy M. Heckman}
\affiliation{Center for Astrophysical Sciences, 
Department of Physics and Astronomy, 
The Johns Hopkins University, 
Baltimore, MD 21218, USA}

\author{Bethan L. James}
\affiliation{Space Telescope Science Institute,
3700 San Martin Drive, 
Baltimore, MD 21218, USA}

\author{Knox S. Long}
\affiliation{Space Telescope Science Institute,
3700 San Martin Drive, 
Baltimore, MD 21218, USA}
\affiliation{Eureka Scientific, 
Inc. 2452 Delmer Street, Suite 100, 
Oakland, CA 94602-3017, USA}

\author{Evan D. Skillman}
\affiliation{Minnesota Institute for Astrophysics, 
School of Physics and Astronomy,
116 Church Street S.E.,
University of Minnesota, 
Minneapolis, MN 55455, USA}

\author{Bradley C. Whitmore}
\affiliation{Space Telescope Science Institute,
3700 San Martin Drive, 
Baltimore, MD 21218, USA}



\begin{abstract}
Stellar populations are powerful tools for investigating the  evolution of extragalactic environments. We present the first UV integrated-light spectroscopic observations for 15 young star clusters in the starburst M83 with a special focus on metallicity measurements. The data were obtained with the Cosmic Origins Spectrograph (COS) onboard the Hubble Space Telescope. We analyse the data applying an abundance technique previously used to study an optical set of star clusters. We estimate a central metallicity of [Z] = $+$0.20 $\pm$ 0.15 dex in agreement with those obtained through independent methods, i.e. $J$-band and blue supergiants.
We estimate a UV metallicity gradient of $-$0.041 $\pm$ 0.022 dex kpc$^{-1}$ consistent with the optical metallicity gradient of $-$0.040 $\pm$ 0.032 dex kpc$^{-1}$ for $R/R_{25}<0.5$. Combining our stellar metallicities, UV and optical, with those from \ion{H}{2} regions (strong-line abundances based on empirical calibrations) we identify two possible breaks in the gradient of M83 at galactocentric distances of $R\sim0.5$ and $1.0\:R_{25}$. If the abundance breaks are genuine, the metallicity gradient of this galaxy follows a \textit{steep-shallow-steep} trend, a scenario predicted by three-dimensional (3D) numerical simulations of disc galaxies. The first break is located near the corotation radius. This first \textit{steep} gradient may have originated by recent star formation episodes and a relatively young bar ($<$1 Gyr). In the numerical simulations the \textit{shallow} gradient is created by the effects of dilution by outflow where low-metallicity material is mixed with enriched gas. And finally, the second break and last \textit{steep} gradient mark the farthest galactocentric distances where the outward flow has penetrated.

\end{abstract}



\section{Introduction}\label{sec:intro}
Measurements of the chemical composition of galaxies are a crucial part of our toolbox for understanding galaxy formation and evolution. Metallicities (e.g. from forbidden, collisionally excited lines in \ion{H}{2} regions, \citealt{men41,pag79,all79}) and abundance gradients in galaxies can provide us with a wealth of information on their past histories. Measurements of the chemical abundances of galaxies can lead to strong constraints on their chemical enrichment including star formation scenarios, star formation rates, gas inflow rates, and Initial Mass Function \citep[IMF, ][]{mcw97}. \par
The metallicities of individual galaxies are primarily moderated by two phenomena: the material processed inside stars, and the exchanges occurring between galaxies and the intergalactic medium (IGM). Studies have shown that there is a tight correlation between the stellar mass of galaxies and their metallicity \citep{leq79, zar94, tre04, mai08, kud12, berg12}. This mass-metallicity relation (MZR) sets crucial observational constraints for models of galaxy evolution which are used to understand the formation of galaxies across cosmic time. Similarly, galactic metallicity gradients are sensitive to complex dynamics such as merging events, and flows of gas from the IGM onto galaxies \citep{pra00, bre09,bre16, kud15, ho15, gri18}.  Given the importance of accurate metallicities, it is crucial to obtain reliable abundance measurements in order to correctly interpret the processes influencing these relations. \par

For several decades, measurements from both metallicity gradients and the MZR of star-forming galaxies have routinely relied on the analysis of emission lines of \ion{H}{2} regions. These studies typically apply two main methodologies: strong-line analysis, and T$_{e}$-based method. The former method is based on the flux ratios of some of the strongest forbidden lines, typically [\ion{O}{2}] and [\ion{O}{3}], with respect to H \citep{pag79,ski89,mcg94}. One of the advantages of this method is that these emission lines are easily detected across a broad range of metallicities with the caveat that the strong-line calibrations have shown large discrepancies in their measurements \citep{kew08, bre09}. The T$_{e}$-based method, on the other hand, is known to provide the most accurate measurements of metallicity relying on the flux ratios of auroral to strong lines of the same ion to infer the electron temperature of the gas \citep{rub94,lee04,sta05}. The T$_{e}$-based analysis depends critically on the physical conditions of the gas, more specifically the electron temperature, and provides a direct measurement of the amount of metals required to cool the gas to the observed electron temperature. Until recently, this method had been challenged in the metal-rich regime where temperature sensitive lines are commonly too weak to be detected \citep{sta05,bre05,gaz15}. However, in the last few years robust, high signal-to-noise spectroscopic observations of \ion{H}{2} regions in high metallicity environments have allowed measurements of temperature-sensitive auroral lines as part of the CHemical Abundances of Spirals \citep[CHAOS,][]{berg15,cro15,cro16} project. For a general overview of these two methodologies we refer the reader to the work by \citet{mar12} and \citet{per17}.

Alternatively, stellar populations and individual stars can be used to determine the metallicity of star-forming galaxies. In the last decade, new techniques have been developed to study the chemical contents of nearby galaxies using blue supergiants (BSGs) as well as red supergiants (RSGs). It has been shown extensively that accurate metallicities can be obtained in agreement with measurements inferred using the T$_{e}$-based method using either BSGs or RSGs \citep{kud12,kud13, kud14, hos14, dav10,gaz14,dav15,gaz15,dav17}. The agreement between the nebular and stellar abundances is of the order of $\sim$0.1 dex.\par
With current telescopes, star clusters at distances of several megaparsecs (Mpc) appear unresolved, but it is possible to measure stellar metallicities through the analysis of their integrated light. Because the integrated-light (IL) spectra of most star clusters are only broadened by a few km s$^{-1}$, high resolution observations ($R = \frac{\lambda}{\Delta\lambda} \sim 20,000-30,000$) can be used to extract large amounts of information. Previous studies have successfully measured detailed abundances and overall metallicities from high resolution spectroscopic observations ($R\sim$30,000) of globular clusters \citep[GC, ][]{mcw02,mcw08,ber05,col09,col11, col12, lar12, lar14, lar17, lar18} and young massive clusters \citep[YMC, ][]{lar06,lar08}.  Furthermore, in \citet{her17, her18a, her18b} we apply the analysis technique developed by \citet[][hereafter L12]{lar12} to less ideal cases using lower-resolution observations ($R\sim 5000-10,000$) and are able to derive detailed abundances and overall metallicities of both GCs and YMCs. \par
Although the work on IL spectroscopy for inferring the chemical composition of stellar populations has shown ample potential, most efforts  have focused on the optical and infrared regimes. Young stellar populations, for instance, when observed in the near-infrared regime are strongly dominated by RSGs, which facilitates the analysis by modelling the whole population as a single RSG \citep[$J$-band method; ][]{dav10, gaz14, lar15}. When studying GCs, on the other hand, one needs to implement full population synthesis techniques to perform a similarly detailed abundance analysis. \citet{lar18b} has recently tested the consistency of the L12 approach by analysing spectroscopic observations covering optical and infrared wavelengths. Their study finds that the L12 technique is well suited for the infrared regime, with a special need for accurate modelling of the coolest stars, given their higher contribution to the IL.  By way of contrast, UV spectral coverage for IL analyses of stellar populations to obtain clues to  their chemical evolution is rather unexplored. In the UV we expect a stronger contribution from hot massive stars, than in any other wavelength range \citep{fan88, meu95}.\par

In this paper we extend the initial metallicity study at optical wavelengths of the nearby \citep[4.9 Mpc, ][]{jac09} face-on star-forming galaxy M83 (NGC 5236) to explore the UV regime. In Table \ref{table:gal} we summarise some essential information of this galaxy. In \citet{her18a} we demonstrate that the L12 technique is an effective tool for  measuring overall metallicities even in high metallicity environments, such as that observed in M83 \citep[ $\lbrack$Z$\rbrack= \log\: $Z/Z$_\odot$ $\sim$ 0.3 dex,][]{bre02, bre05, gaz14, dav17}. Uncertainties in the overall metallicities obtained using the L12 method on intermediate-resolution ($R<8,800$) observations of YMCs in M83 are of the order of $\sim$0.1 dex. Furthermore, we find excellent agreement between our stellar metallicities and those inferred through independent methods (i.e., BSGs and $J$-band). We aim to extend our earlier work by adding 15 YMCs to our sample, while testing the feasibility of the  IL analysis when applied to UV spectroscopic observations.  \par
Here we present the IL analysis of 15 YMCs ($<$20 Myr) in M83. Using spectroscopic observations taken with the Hubble Space Telescope (HST) Cosmic Origins Spectrograph (COS), we measure metallicities for each of the clusters. This manuscript is structured as follows: In Section \ref{sec:obs} we describe the observations and the steps applied to reduce the spectroscopic data. In Section \ref{ana} we introduce our analysis technique in detail. Section \ref{results} presents our main results. In Sections \ref{discuss} and \ref{con} we discuss our findings and present our conclusions, respectively. 

%__________________________________________________________________

\section{Observations and Data Reduction}\label{sec:obs}
The work presented here makes use of the data observed as part of HST program ID 14681 (PI: Aloisi) executed in May and June of 2017. The YMCs were acquired applying a standard peak-up target acquisition (TA) in the NUV imaging mode, and observed with the G130M/1291 and G160M/1623 settings providing a wavelength coverage between 1130 and 1800 \r{A} and wavelength-dependent resolution $R\sim 15,000$-20,000. \par
\begin{figure}
   \resizebox{\hsize}{!}
            {\includegraphics[width=11.2cm]{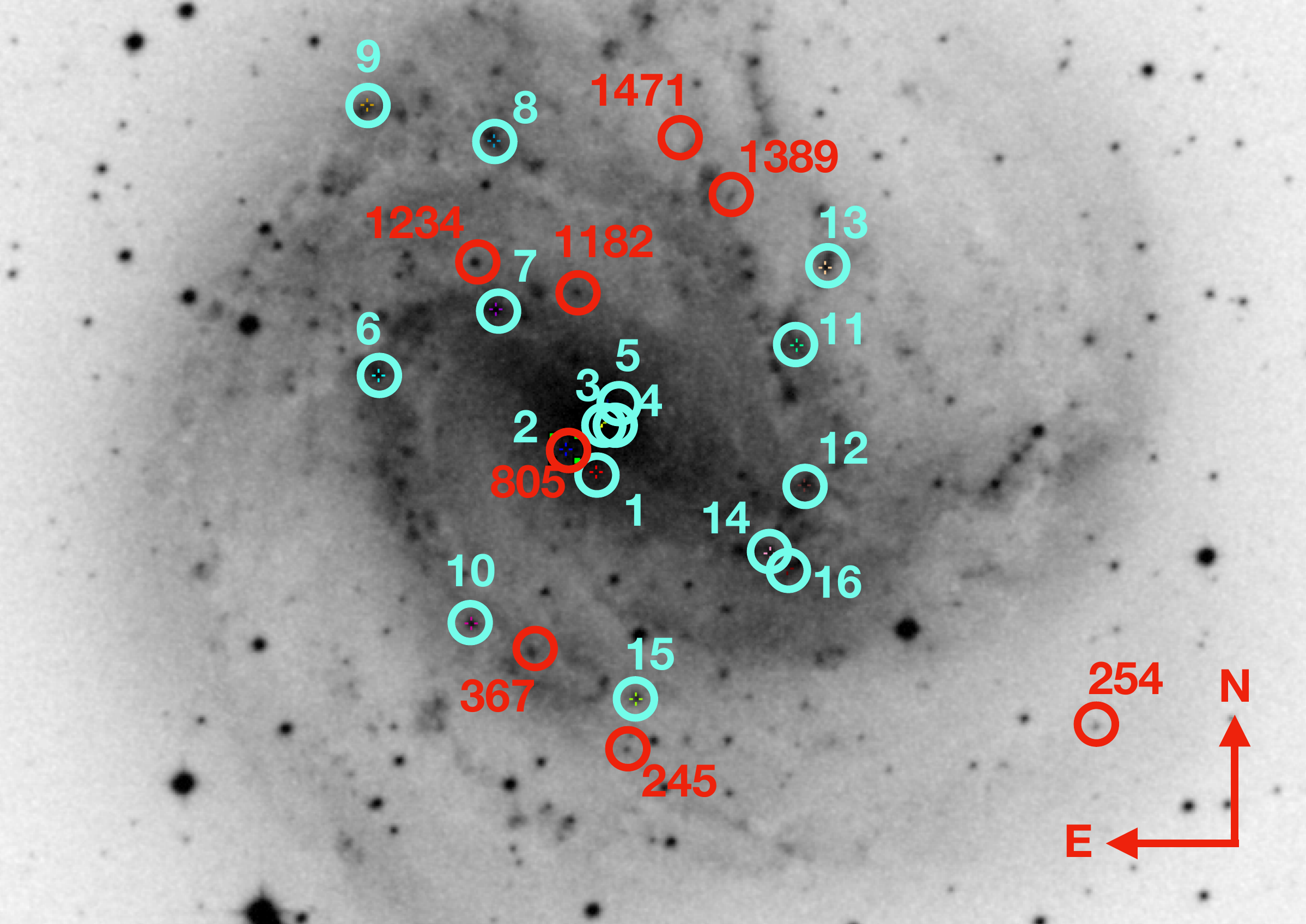}}
      \caption{$R$-band image of M83 taken as part of the Digitized Sky Survey. Red circles locate the clusters studied in \citet{her18a}. Cyan circles show the clusters studied here.}
         \label{Fig:galaxy}
   \end{figure}

The targets studied here were selected from published and unpublished lists of young star clusters observed with the HST Wide Field Camera 3 (WFC3) using both broad- and narrow-band filters as part of the WFC3 Early Release Science Cycle 17 GO/DD program 11360 (PI: O'Connell) and Cycle 19 GO program 12513 (PI: Blair). Observational details of these data are provided in \citet[][Table 1]{bla14}. Our primary selection criterion requires clusters to have magnitudes $m_{\rm F336W}\lesssim 17$. In general, this criterion selected the most massive and youngest clusters and provided a complete magnitude-limited sample. We do not expect any significant bias introduced by the selection on luminosity given that these young clusters are below the limit where mass-metallicity relations begin to appear, e.g. in old GC systems. The selected sample is representative of the present-day composition of M83. In Figure \ref{Fig:galaxy} we mark with cyan circles the YMCs studied as part of this work. These targets cover a range of ages from $\sim$5 to $\sim$20 Myr. The age estimates were calculated following the techniques of \citet{cha10} and \citet{whi11}. The uncertainties in the age estimates using the techniques by \citeauthor{cha10} and \citeauthor{whi11} are of the order of 50\%. In Table \ref{table:obs} we list the individual YMCs, their coordinates, magnitudes, ages, galactocentric distance normalised to the isophotal radius \citep[$R_{25} = 6.44$\arcmin, ][]{dev91}, exposure times for each of the gratings, and their corresponding signal-to-noise (S/N). The S/N values in the last two columns of Table \ref{table:obs} were estimated at wavelengths of 1310 \r{A} and 1700 \r{A} for gratings G130M and G160M, respectively. The S/N for each observation is sufficient to reach the original goal of the program,  which included the detection and measurement of the weakest interstellar lines (e.g., \ion{S}{2} at 1250 and 1253 \r{A}, and \ion{Mg}{2} at 1239 and 1240 \r{A}). \par
The individual spectra were downloaded from the Mikulski Archive for Space Telescopes (MAST) and reduced using the HST calibration pipeline, \textsc{CALCOS v3.3.4} \citep{fox15}. For those cases where more than one association file was recorded per configuration, we take the individual x1d files and further analyze them using the IDL code by the COS Guaranteed Time Observer (GTO) Team \citep{dan10}. Using this software we weight combine the different FP-POS exposures (x1d files) interpolating onto a common wavelength vector accounting for non-Poissonian noise as detailed by \citet{kee12}. After the standard calibration is finished, we bin the final spectra by a COS resolution element (1 resel = 6 pixels) corresponding to the nominal point-spread function. We check the wavelength calibration accuracy by comparing the laboratory wavelengths from MW absorption lines, i.e. Si II 1190.42 \r{A}, against our measured wavelengths. We find offsets of the order of $\sim$ 0.05 \r{A} within the wavelength accuracy of COS (0.06 \r{A}). \par

\begin{table}
\caption{M83: Galaxy Parameters}
\label{table:gal}
\centering 
\begin{tabular}{lc}
\hline \hline
Parameter & Value \\
\hline\\
R. A. (J2000.0) & 204.253958\\
Dec (J2000.0) & -29.865417\\
Distance$^{a}$ & 4.9 Mpc\\
Morphological type & SAB(s)c \\
$R_{25}\:^{b}$ & 6.44$\arcmin\;$  (9.18 kpc)\\
Inclination$^{b}$ & 24\degr\\
Position Angle$^{c}$ & 45\degr\\
Heliocentric radial velocity & 512.95 km s$^{-1}$\\
\hline
\end{tabular}
\begin{minipage}{15cm}~\\
\textbf{Notes.} All parameters from the NASA Extragalactic \\Database (NED), except where noted.\\
 \textsuperscript{$a$}{\citep{jac09}} \\
 \textsuperscript{$b$}{\citep{dev91}} \\
 \textsuperscript{$c$}{\citep{com81}}\\
 \end{minipage}
\end{table}
   
\begin{table*}
\caption{Properties of the young massive clusters and their COS Observations}
\label{table:obs}
\centering 
\begin{tabular}{ccccccccccccc}
\hline \hline
Cluster & RA & Dec & $m_{\rm F336W}$ & Age & $R/R_{25}$ $^{a}$ & \multicolumn{2}{c}{$t_{\rm exp}$ (s)} & \multicolumn{2}{c}{S/N $^b$ (resel$^{-1}$)}   \\
 & (J2000) & (J2000) & (mag) & (Myr) & & G130M & G160M &  G130M & G160M\\
\hline\\
M83-1 & 204.2527583&-29.8739111 & 17.0 & 10 $\pm$ 5 & 0.08 &2368.032 & 8114.336  & 1.5 & 1.3  \\
M83-2 & 204.2576792&-29.8703833 & 16.14 & 10 $\pm$ 5 & 0.06 &2120.032 & 5652.448  & 2.2 & 1.5 \\
M83-3 & 204.2517375&-29.8666222 & 15.10 & 5 $\pm$ 3 & 0.02 &400.064 & 1304.032  & 6.9 & 5.5 \\
M83-4 & 204.2514333&-29.8662056 & 14.40 & 10 $\pm$ 5 & 0.02 & 500.032 & 1208.000  & 3.1 & 2.2 \\
M83-5 &204.2504500&-29.8642500 & 14.85 & 20 $\pm$ 10 & 0.04 & 1599.872 & 3072.384  & 0.9  & 0.6\\
M83-6 & 204.2895625&-29.8588083 & 16.46 & 6 $\pm$ 3 & 0.34 & 1036.032 & 3644.672  & 4.0 & 3.5 \\
M83-7 & 204.2692667&-29.8495917 & 14.79 & 4 $\pm$ 2 & 0.21 &539.968 & 1135.968 & 2.2 & 1.3\\
M83-8 & 204.2688667&-29.8246056 & 16.76 & 10 $\pm$ 5 & 0.41 &2986.752 & 7182.528  & 6.8 & 4.1 \\
M83-9 & 204.2904083&-29.8187833 & 15.65 & 10 $\pm$ 5 & 0.56 &1056.064 & 3616.640  & 3.6  & 2.5\\
M83-10 & 204.2746000&-29.8959056 & 16.94 & 5 $\pm$ 3 & 0.34 &1780.000 & 5834.624  & 8.0  & 5.5 \\
M83-11 & 204.2179875&-29.8557389 & 16.65 & 7 $\pm$ 3 & 0.35 & 2091.936 & 5652.672  & 5.4 & 3.2 \\
M83-12 & 204.2171125&-29.8764111 & 16.84 & 7 $\pm$ 3 & 0.36 &2891.744 & 7276.544  & 6.3  & 4.3\\
M83-13 & 204.2127208&-29.8444639 & 16.76 & 4 $\pm$ 2 & 0.43 & 2232.064 & 5652.672  & 5.6  & 3.4\\
M83-14 & 204.2230583&-29.8863750 & 15.30 & 5 $\pm$ 3 & 0.35 &400.032 & 1295.936  & 2.8 & 1.7 \\
M83-15 & 204.2465708&-29.9075222 & 16.24 & 5 $\pm$ 3 & 0.40 &1040.000 & 3644.384  & 2.1 & 1.6 \\
M83-16 & 204.2204833&-29.8887472 & 16.79 & 5 $\pm$ 3 & 0.38 &2987.840 & 7168.512  & 5.4 & 3.4\\
\hline
\end{tabular}
\begin{minipage}{15cm}~\\
 \textsuperscript{$a$}{Calculated adopting the parameters listed in Table \ref{table:gal}.}\\
  \textsuperscript{$b$}{Estimated at wavelengths of 1310 \r{A} and 1700 \r{A} for G130M and G160M, respectively.}\\
 \end{minipage}
\end{table*}

%__________________________________________________________________

\section{Abundance Analysis}\label{ana}
The work presented here uses the IL analysis tool developed by L12. The full spectral fitting technique by L12 was first introduced in a study of extragalactic GCs in the Fornax dwarf galaxy requiring high-resolution observations (R$\sim$30,000), and continued being applied and tested in other Galactic and extragalactic stellar populations \citep{lar14, lar17, lar18}. In a recent publication, \citet{her18a} showed that this same tool can be successfully used to measure overall metallicities of young stellar populations in intermediate-resolution (R$<$8,800) observations covering optical wavelengths. \par
Briefly summarised, the L12 technique combines information from the Hertzsprung-Russell diagram (HRD), stellar atmospheric models, and synthetic spectra. We encapsulate the L12 analysis in four main steps: 1) Derive stellar parameters of individual stars in the stellar population in question from their colour magnitude diagrams (CMDs) or theoretical isochrones. For the work presented here we make use of theoretical isochrones with a selection based on the ages found in the literature (see Table \ref{table:obs}) and the mean metallicity of the galaxy \citep{bre02}. 2) Produce stellar atmospheric models with said stellar parameters. 3) Create synthetic spectra for all of the stars present in the clusters and combine them into a single synthetic IL model spectrum. 4) Compare the synthetic spectrum to the observations. Steps 3-4 are repeated iteratively modifying the individual abundances until the best match to the empirical data is obtained. We apply a golden-section search technique looking for a minimum between the metallicity range $-$1.0 $<$ [Z] $<$ $+$0.8 dex. Since the overall metallicity of M83 is expected to be around [Z] $\sim$ 0.3 dex, we select limits bracketing any reasonable expectation of what the range may be.\par

\subsection{Derivation of Stellar Parameters}\label{sec:stel_par}
Similar to the approach in \citet{her18a}, in the first step of the analysis we generate a Hertzsprung-Russell diagram (HRD) covering every evolutionary stage expected in the individual stellar populations. We create the HRDs using stellar isochrones from \textsc{PARSEC v.1.2.S} \citep{bre12}. The isochrone grid covers metallicities between $-$2.2 $<$ [Z]$<$ $+$0.8 dex. The theoretical models are initially selected assuming a metallicity of [Z] = $+$0.3 dex for M83 \citep{bre02} and adopting the ages listed in Table \ref{table:obs}. \par
From the isochrones we directly extract the stellar masses ($M$), effective temperatures (T$_{\rm eff}$), and the surface gravities (log $g$). The stellar radii are estimated assuming a gravitational acceleration at the surface of the star of 

\begin{eqnarray}
g = \frac{G M}{R^2} 
\end{eqnarray}

Using the Sun as fiducial star, we express the stellar radii in solar units. Said parameters ($M$, T$_{\rm eff}$, log $g$, $R$) are used applying a lower mass limit of 0.4 $M_{\odot}$ and a \citet{sal55} exponent of $\alpha =2.35$ for an IMF following a power law:
\begin{eqnarray}
\frac{dN}{dM} \propto M^{- \alpha}
\end{eqnarray} 
Following the same recipe applied to the M83 YMCs by \citet{her18a}, we use the following microturbulent velocity values, $v_{t}$: for stars with T$_{\rm eff}<$6000 K we adopt a value of $v_{t}$= 2 km s$^{-1}$, for stars with T$_{\rm eff}$ between 6000--22000 K we adopt a value of $v_{t}$= 4 km s$^{-1}$ \citep{lyu04}, and for stars with T$_{\rm eff}>$ 22000 K we use $v_{t}$= 8 km s$^{-1}$. \par

   \begin{figure*}
   %\resizebox{\hsize}{!}
            {\includegraphics[scale=0.5]{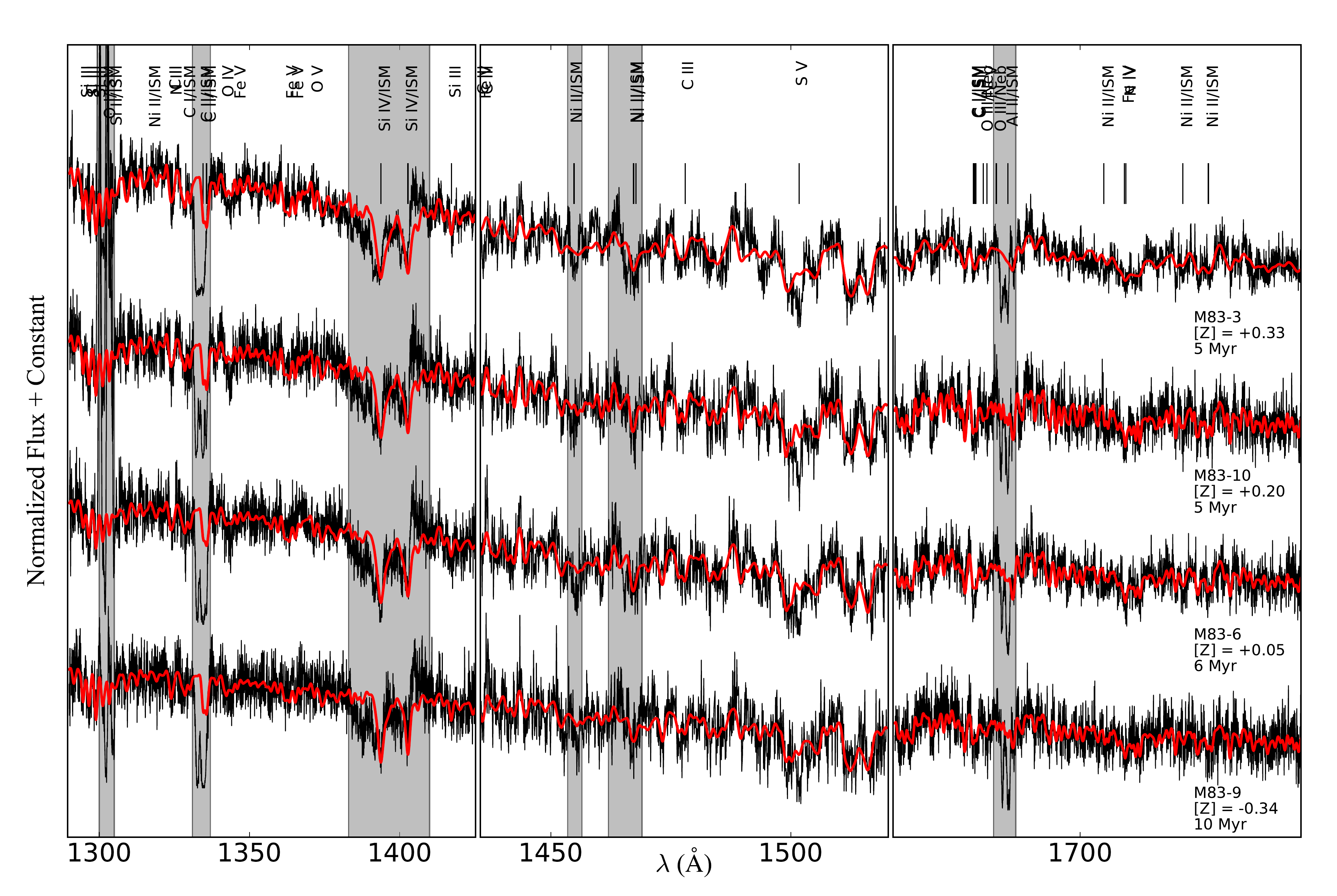}}
      \caption{Normalized spectra for a selected sample of YMCs in M83. We normalized the spectra to their flux at 1290\r{A} for visual effects. The COS observations are shown in black. Best model fits are displayed in red. The individual cluster names are shown below their corresponding spectra as well as their overall metallicities and ages. We display the clusters from bottom to top in order of increasing metallicity. The grey regions correspond to the wavelength coverage contaminated by strong geocoronal, stellar winds and ISM lines. We exclude these regions from the fitting procedure.}
         \label{Fig:spec_fits}
   \end{figure*}

\subsection{Stellar Atmospheric Models}\label{sec:atm_models}
The initial L12 technique uses model atmospheres computed by the publicly distributed software \textsc{ATLAS9} \citep{kur70}. \textsc{ATLAS9} produces local thermodynamic equilibrium (LTE) one-dimensional (1D) plane-parallel models which are slightly less ideal for the coolest giants expected to be present in younger populations and in more metal-rich clusters. In previous publications where we use the L12 technique to explore younger populations using optical data \citep{her17,her18a}, we combine \textsc{ATLAS9} models for stars with T$_{\rm eff} >$5000 K with \textsc{MARCS} stellar atmospheres \citep{gus08} for those stars with cooler temperatures, T$_{\rm eff}< $5000 K. In general, no single stellar type dominates the IL observations in the optical wavelength regime. UV studies of stellar populations, on the other hand, are dominated by hot stars. According to \citet{meu95}, for a stellar population of $\sim$10 Myr and a standard IMF, 50\% of the UV light is produced by stars with masses $<$28 M$_\odot$ and only 5\% by stars with masses $<$8 M$_\odot$. Stars with masses $<$ 5 M$_\odot$ produce negligible emission in the UV \citep{meu95}. \par
\begin{figure}
   \centering
   %\resizebox{\hsize}{!}
            {\includegraphics[scale=0.39]{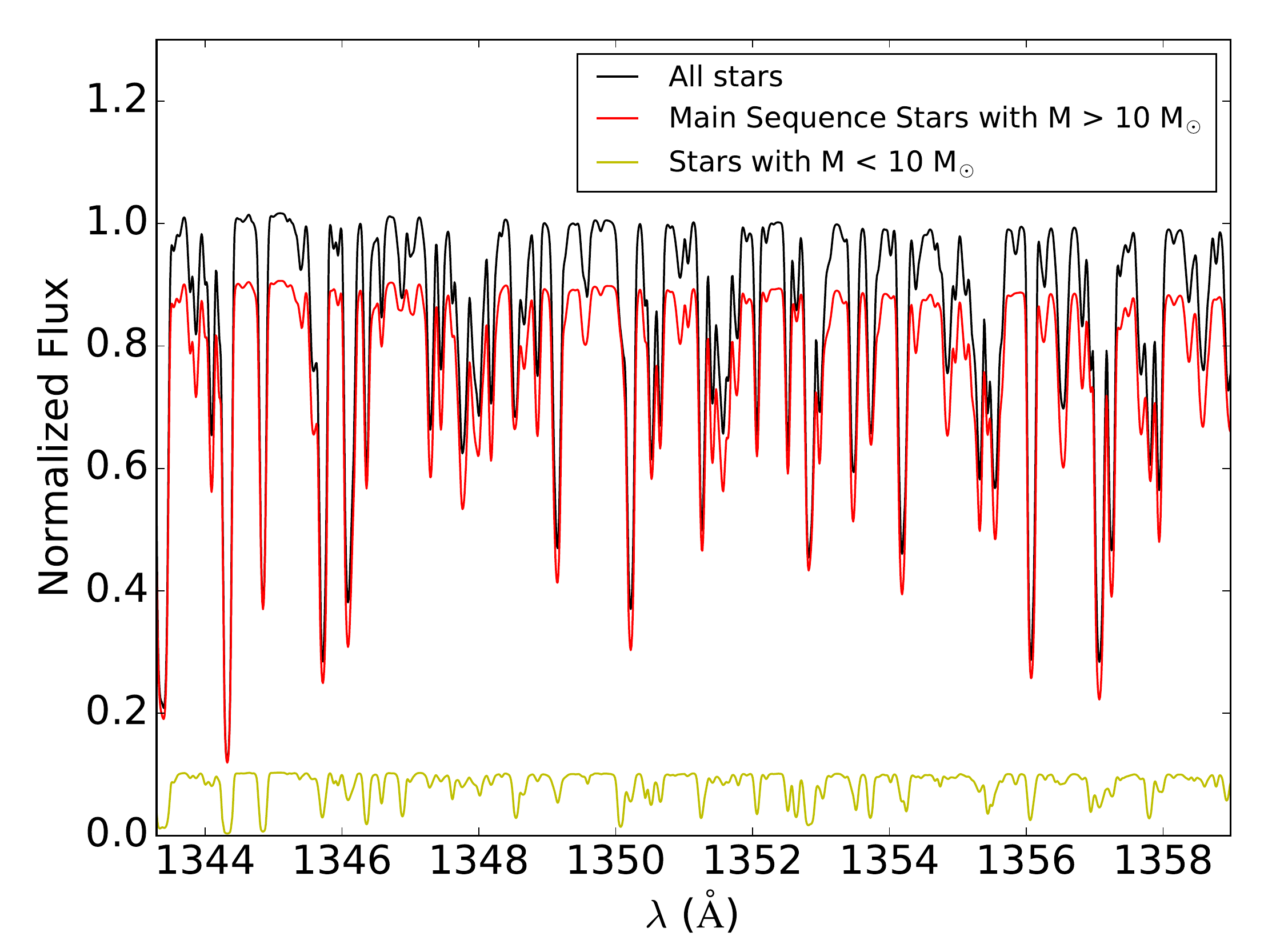}}
      \caption{Contribution to the integrated light for all stars (in black), massive stars ($>$ 10 M$_{\odot}$, in red), and stars with $<$10 M$_{\odot}$ (including giants, in yellow) in the FUV for a population of stars with an age of 7 Myr, and solar metallicity. }
         \label{fig:hot_stars}
   \end{figure}

To confirm the expected contribution from the most massive stars to the integrated-light spectrum of these young populations, we generate model spectra for a stellar population with an age of 7 Myr and solar metallicity, [Z] = 0.0. In Figure \ref{fig:hot_stars} we show the full normalised integrated-light spectrum (in black), and compare that to the contribution from the most massive ($>$ 10 M$_{\odot}$) and hottest stars ($>$ 10,000 K, in red). We also include in yellow, the contribution from stars with lower masses and temperatures (including giants). We can see that in the FUV, massive stars contribute $\sim$90 \% of the total flux for a stellar population of 7 Myr. Given that in the UV regime the contribution from cool stars is minimised \citep{zin07}, our analysis here uses atmospheric models from \textsc{ATLAS9} only.   \par
We note that the analysis done here is entirely based on LTE modelling, without implementing any corrections for non-LTE (NLTE) effects. For IL observations such adjustments are more complex than those applied to individual stars given that NLTE corrections are dependent on the stellar type (log $g$, T$_{\rm eff}$), as well as their metallicity. For reference, NLTE corrections can reach values as high as 0.4 dex. In general, NLTE corrections for individual stars for [Fe/H]  have been found to be around $\lesssim$0.1 dex \citep{ber12}, for some $\alpha$-elements are expected to be between 0.1 and 0.4 dex \citep{ber15}, and for some light-elements between 0.2 and 0.3 dex \citep{gra99}. We note that these corrections apply to abundance analysis of cooler stars ($\lesssim 5000$ K). To provide a more quantitative statement on the NLTE corrections applicable to the IL analysis performed here, we would require a detailed modeling of the NLTE effects.\par

\subsection{Synthetic Spectra}
We use the atmospheric models described in Section \ref{sec:atm_models} to create synthetic spectra. We generate one spectrum for each stellar type, based on the model atmospheres. Since we only make use of \textsc{ATLAS9} models we only require the use of \textsc{SYNTHE} \citep{kur79, kur81} when producing the synthetic spectra. We generate these models, both atmospheric and spectral, ``on-the-fly" allowing us to obtain models with the desire stellar parameters, i.e. with no dependancies on pre-computed spectral libraries. The spectral synthesis code uses line lists published by \citet{cas04}. The spectra are created at relatively high resolutions, $R\sim$500,000, and combined onto a single synthetic IL spectrum. \par
The current software allows for an additional step of estimating the best-fitting Gaussian dispersion value, $\sigma_{\rm sm}$, used to smooth the model and obtain synthetic spectra with comparable resolutions to those of the COS observations. We note that for this analysis we match the resolution of the observations alone; the L12 technique does not require matching the S/N of the synthetic spectra to the observations. As an initial run, we fit the COS data allowing the $\sigma_{\rm sm}$ and overall metallicity, [Z], to vary. This first calculation provides us with the best smoothing parameter to be used in the following iterations. Our estimates of $\sigma_{\rm sm}$ should account for the instrumental resolution, internal velocity dispersions in the clusters and stellar rotation and macroturbulence. In Table \ref{table:derived_vel} we list our final values of $\sigma_{\rm sm}$. In previous publications where the analysis focused on the optical part of the IL spectra of stellar populations, we estimate the line-of-sight velocity dispersions ($\sigma_{\rm 1D}$) for each star cluster assuming that the instrumental resolution ($\sigma_{\rm inst}$) is set solely by the slit width, and that the main components in defining the $\sigma_{\rm sm}$ are the $\sigma_{\rm 1D}$ and the $\sigma_{\rm inst}$ following
\begin{equation}
\sigma_{\rm sm}^{2} = \sigma_{\rm inst}^{2} + \sigma_{\rm 1D}^{2}
\end{equation}
Here, however, we analyse the UV regime of the IL of young star clusters. As mentioned above, the IL of young stellar populations at UV wavelengths is dominated by hot massive stars \citep{con13}. At ages of only a few Myr we expect massive stars, O- and B-type (stellar types for which binarity is non-negligible \citealt{van98, san12}), to highly contribute to the light in the integrated spectrum. These stellar types have been observed to have rotational velocities of $\sim$80 km s$^{-1}$ or higher \citep{hun08,ram13}. This means that such velocities could in principle be imprinted in the broadening of the stellar lines. This assumption agrees with the observed $\sigma_{\rm sm}$ values, where we find remarkably higher values when compared to what is expected by accounting \textit{only} for the instrumental resolution ($\sigma_{\rm inst}$) and the cluster velocity dispersion ($\sigma_{\rm 1D}$). Although accurate measurements of the rotational velocities of such stars in these young populations are of general interest, they are beyond the scope of this paper. Furthermore, COS is optimised for point sources using an aperture rather than a slit, meaning that the instrumental resolution, $\sigma_{\rm inst}$, is strongly dependent on the extension of the sources. We note however that an accurate assessment of the instrumental resolution of our observations is outside of the main goals of this work. \par

\subsection{Synthetic spectra vs. COS observations}
At this stage the smoothed synthetic IL spectrum is directly compared to the COS data. This comparison is done normalizing the observed spectra applying either a cubic spline with three knots (for wavelength bins $>$100 \r{A}) or a first order polynomial (for wavelength bins $<$100 \r{A}) to match the continua of the model and COS observations. In this process the entire spectrum, within the established wavelength windows, is fit at once. The code fits continuum and absorption features simultaneously. The L12 software also allows the user to define weights for each of the pixels in the spectrum, where pixels with weights of 0.0 are entirely excluded. Any spectral regions with weights set to 0.0 are not used in the comparison between the synthetic spectrum and the observation. For the analysis presented here we set the weights to 0.0 in wavelength regions with strong geocoronal, stellar winds, and ISM contamination (see Figure \ref{Fig:spec_fits}).\par
The steps described above are repeated iteratively, modifying each time the abundances until the best match is obtained through the minimisation of the $\chi^{2}$. We note that the metallicities of the isochrones are chosen to self-consistently match those derived from the abundance analysis, which in some cases requires several iterations before converging to the final metallicities. The overall metallicities derived in this work, [Z], are measurements of the integrated abundances of different chemical elements (i.e. $\alpha$, light, Fe-peak). This analysis assumes that the metals have a relative solar abundance pattern, meaning that the metallicities obtained here are overall scalings of all of the abundances (relative to Solar). This is one of the limitations imposed primarily by the S/N of our observations. Here we use the Solar composition as published by \citet{gre98}. 

%
%________________________________________________________________
\section{Results}\label{results}
The steps described in Section \ref{ana} are followed once we remove the radial velocity component (second column in Table \ref{table:derived_vel}) from the  COS observations. We fit for the overall metallicity, [Z], keeping the $\sigma_{\rm sm}$ fixed (third column in Table \ref{table:derived_vel}). Given that  \textsc{ATLAS9} does not model any spectral features originating outside of the stellar photospheres (i.e. ISM lines), we avoid wavelength regions strongly affected by geocoronal/stellar-wind/ISM contamination. We limit the metallicity analysis of the stellar populations to three wavelength bins, 1290--1430 \r{A}, 1435--1520 \r{A} and 1625--1790 \r{A} masking geocoronal/ISM lines within these ranges. These windows exclude the strong geocoronal Lyman $\alpha$ emission line, as well as the strongest ISM absorption lines \citep{lei11} dominating the wavelength coverage falling on Segment B of the G130M setting. \par

We assume a metallicity of [Z]=$+$0.3 in the initial selection of the input isochrone. In \citet{her17}, for the NGC1705-1 YMC, we find that using an input isochrone with the initially inferred metallicity does not provide the best model spectrum. As part of the analysis performed here we inspect the final fits and find that these model spectra agree well with the observations.  We observe that the metallicity of the isochrone used in the last iteration consistently converges onto the inferred metallicity of the YMC in question. Furthermore, in \citet{her17} we test the sensitivity of our results to the assumed parameters, [Z] and age. We find that varying the age ($+$25 Myr) and metallicity ($+$0.15 dex) of the initial input isochrone have a minor effect on abundances such as [Fe/H], on the order of $\lesssim$0.1 dex. The individual bins, their final metallicities and their 1-$\sigma$ uncertainties are listed in Table \ref{tab:1}. The 1-$\sigma$ uncertainties are calculated by varying the metallicity of the models until $\chi^{2} = \chi_{min}^{2} + 1$. \par
In Figure \ref{Fig:spec_fits} we show normalised synthesis fits for a selected number of YMCs in our sample. The grey regions correspond to the wavelength coverage affected by strong geocoronal, stellar winds and ISM lines; these regions are excluded from the fitting procedure. The final weighted averaged metallicities are listed in the fifth column of Table \ref{table:derived_vel}. Based on the comprehensive list of stellar photosphere lines by \citet{lei11}, we point out that our inferred metallicities are primarily sensitive to C, Si, N, O, and Fe species. Since the scatter in the individual measurements is larger than the errors obtained from the $\chi^2$ fitting, we estimate the final uncertainties in the metallicities listed in Table \ref{table:derived_vel} accounting for the number of individual bins ($N=3$) and their standard deviation following 
\begin{eqnarray}
\sigma_{\rm err} = \frac{\sigma_{\rm STD}}{\sqrt{N -1}}
\end{eqnarray}
We find that the S/N of the M83-5 observations is too low to use the L12 technique. For this target specifically, we are not able to obtain reliable measurements in any of the three bins. 

\begin{table}
\caption{Derived parameters of YMCs in M83. }
\label{table:derived_vel}
\centering 
\begin{tabular}{ccccc}
\hline \hline
Cluster & $v_{\rm rv}$  & $\sigma_{\rm sm}$ & $v_{\rm residual}$ & [Z] \\
 & (km s$^{-1}$) & (km s$^{-1}$) & (km s$^{-1}$) &(dex)\\
\hline\\
M83-1 & 468 $\pm$ 43& 94 $\pm$ 15 & $-$69 &$+$0.57 $\pm$ 0.10\\
M83-2 & 503 $\pm$ 8& 66 $\pm$ 5 & $-$14 &$-$0.14 $\pm$ 0.17\\
M83-3 & 508 $\pm$ 9& 104 $\pm$ 21 & $-$12&$+$0.33 $\pm$ 0.15\\
M83-4 & 530 $\pm$ 3& 88 $\pm$ 16 &  13 &$+$0.02 $\pm$ 0.16\\
M83-5 &  -- & -- & -- & -- \\
M83-6 & 443 $\pm$ 7 & 93 $\pm$ 5 &  $-$23 &$+$0.05 $\pm$ 0.11\\
M83-7 & 455 $\pm$ 17& 89 $\pm$ 1 & $-$10 &$+$0.21 $\pm$ 0.18\\
M83-8 & 464 $\pm$ 12& 82 $\pm$ 8 &  4 &$-$0.04 $\pm$ 0.14\\
M83-9 & 407 $\pm$ 22& 87 $\pm$ 19 & $-$52 &$-$0.34 $\pm$ 0.08\\
M83-10 & 514 $\pm$ 1& 78 $\pm$ 9 & $-$14 &$+$0.20 $\pm$ 0.15\\
M83-11 & 524 $\pm$ 22& 92 $\pm$ 10 & $-$16  &$-$0.03 $\pm$ 0.09\\
M83-12 & 539 $\pm$ 13& 88 $\pm$ 13 & $-$29 &$+$0.12 $\pm$ 0.14\\
M83-13 & 517 $\pm$ 1& 85 $\pm$ 15 & $-$10 &$+$0.06 $\pm$ 0.13\\
M83-14 & 560 $\pm$ 15& 75 $\pm$ 8 & $-$14 &$+$0.12 $\pm$ 0.19\\
M83-15 & 570 $\pm$ 4& 83 $\pm$ 27 & 11 &$-$0.07 $\pm$ 0.08\\
M83-16 & 539 $\pm$ 4& 90 $\pm$ 2 & $-$37 &$+$0.18 $\pm$ 0.14\\
\hline
\end{tabular}
\end{table}

\subsection{Input assumptions}\label{sec:input}
The accuracy of the L12 technique amply relies on the input accuracy. In this section we explore how sensitive our metallicity estimates are to the selection of theoretical models, more specifically the choice of IMF and the input isochrone ages.\par
As described in Section \ref{sec:stel_par}, after the selection of the theoretical isochrones we apply a mass limit following a Salpeter IMF. To test the dependence of our inferred metallicities to the choice of IMF we recalculate our abundances applying a Kroupa IMF instead \citep{kro01}. In the last column of Table \ref{table:input} we list our results as the difference from our nominal values (listed in Table \ref{table:derived_vel}). Since the UV integrated-light spectra of these YMCs is dominated by the most massive stars, and given that the main differences between the Salpeter and Kroupa IMFs are found in the low-mass regime, changing the choice of IMF modifies our metallicities on average by $<<$ 0.1 dex.\par
\citet{her18a} also investigate the intrinsic dependence of the metallicity analysis in the optical regime on the selection of models, more precisely the upper limit uncertainties, $+\sigma_{\rm Age}$. Varying their adopted ages by a factor of 2, they observe on average changes in their metallicity estimates of the order of $\sim$0.1 dex. We note that the average YMC age in the \citeauthor{her18a} sample is $\sim$65 Myr, compared to the average age of the sample analysed here of $\sim$ 7 Myr. In an effort to repeat a similar test as that described in \citet{her18a} we modify the input ages by $\pm$1$\sigma$, as listed in Table \ref{table:obs}. We list our results in columns 2 and 3 of Table \ref{table:input}. \par
We find that increasing the age of the input isochrone by $+\sigma_{\rm Age}$ modifies the metallicity estimates on average by 0.199 dex, higher than the average change observed in the optical measurements. If we instead change the age of the input model by $-\sigma_{\rm Age}$ we find that the inferred metallicities change on average by 0.268 dex. The observed metallicity scatter in the different bins is consistently higher when decreasing the ages of the input isochrones, $-\sigma_{\rm Age}$, this compared to those from the nominal ages or the older models, $+\sigma_{\rm Age}$. The high sensitivity observed at the youngest ages, $-\sigma_{\rm Age}$, is due to the high activity in the clusters. Large evolutionary changes, primarily originating from the most massive stars, are experienced by young clusters in the ages between 2 and 10 Myr \citep{por10}. 

\begin{table}
\caption{Sensitivity to isochrone selection and IMF. }
\label{table:input}
\centering 
\begin{tabular}{cccc}
\hline \hline
Cluster & \multicolumn{2}{c}{$\Delta$ Age}  &IMF/Kroupa \\
 &   $-\sigma_{\rm Age}$  & $+\sigma_{\rm Age}$   & \\
\hline\\
M83-1 & $-$0.431 & $+$0.069 & $+$0.045\\
M83-2 &$-$0.176 & $-$0.019& $-$0.009\\
M83-3 & $-$0.141&  $-$0.454 & $-$0.020\\
M83-4 & $+$0.203& $-$0.124 &  $-$0.001\\
M83-5 &  -- & -- & -- \\
M83-6 & $+$0.358 & $-$0.156 &  $+$0.007\\
M83-7 & $+$0.307 & $-$0.197 &  $+$0.001\\
M83-8 & $-$0.299 & $-$0.159 &  $+$0.003\\
M83-9 & $+$0.292 & $-$0.228 &  $-$0.002\\
M83-10 & $+$0.287 & $-$0.271 & $+$0.007 \\
M83-11 & $+$0.201& $-$0.145 &  $-$0.001 \\
M83-12 & $-$0.188 & $+$0.281 & $+$0.001\\
M83-13 & $+$0.290& $-$0.177 &  $+$0.003\\
M83-14 & $+$0.226 & $-$0.280 &  $+$0.005\\
M83-15 & $+$0.409 & $-$0.255 &  $-$0.001\\
M83-16 & $+$0.219 & $-$0.175 &  $-$0.004\\
\hline
\end{tabular}
\end{table}

\subsection{Vertical velocity dispersions}\label{sec:vel}
The radial velocities ($v_{\rm rv}$) of star clusters rotating in the disc of M83 are described by 
\begin{equation}
 v_{\rm rv}\: = \: v_{\phi} \: \cos\: \phi\: \sin\: i\; +\; v_{R}\: \sin\: \phi \: \sin\: i \; +\;
 v_{z}\: \cos\: i \;+\; v_{\rm sys}
 \end{equation}
where $v_{\phi}$, $v_{R}$ and $v_{z}$ are the rotational/azimuthal, radial and vertical velocity components, $v_{\rm sys}$ is the systemic velocity of the galaxy (512.95 km s$^{-1}$ from Table \ref{table:gal}), $\phi$ is the position angle from the principal axis in the galaxy plane and $i$ is the inclination angle. As per convention we measure $\phi$ counter clockwise from the receding side of the galaxy \citep{teu02}. We also note that even though M83 is nearly face-on it is important to take into account the effect of rotation as this is still considerable. \citet[][and references therein]{mor97} have observed that the galactic gas and the stellar component might be kinematically linked. They find that some galaxies exhibiting warps in their gas, often display stellar warps as well. Furthermore, \citet{mor94} find that in NGC 5907 both the stellar and gas warps have similar orientations. In the work presented here, the rotational component is removed from our radial velocities, $v_{\rm rv}$, by subtracting $v_{\phi} \: \cos\: \phi\: \sin\: i\;$ where $v_{\phi}$ is taken from the study of molecular gas in M83 by \citet{lun04}.\par
\begin{figure}
   %\resizebox{\hsize}{!}
   \centering
            {\includegraphics[scale=0.54]{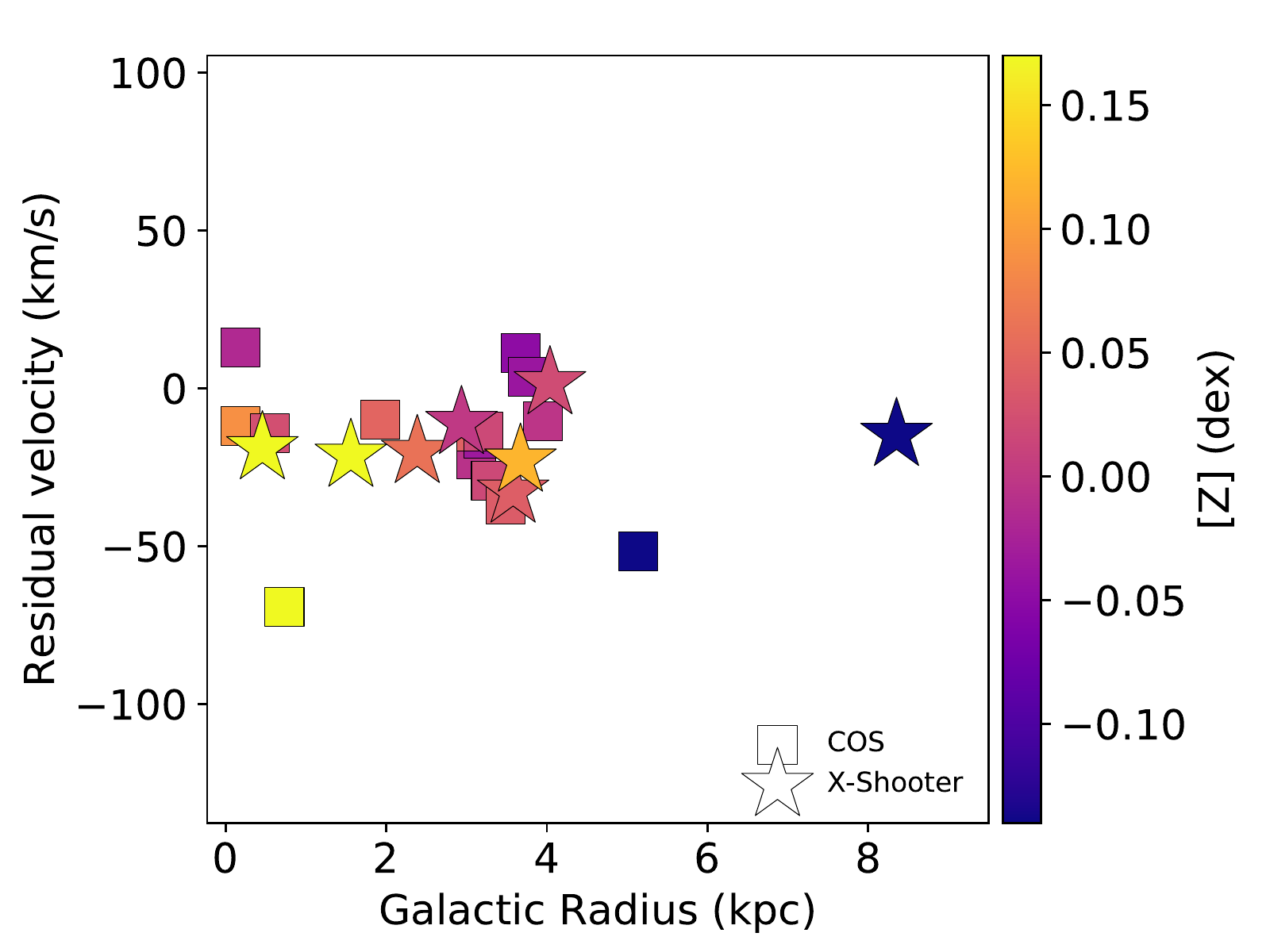}}
      \caption{Cluster velocities after the removal of rotational ($v_{\phi}$) and systemic velocities ($v_{\rm sys}$) as a function of galactocentric distance colour coded by metallicity. Square and star symbols display the measurements from the COS and  X-Shooter observations, respectively.}
         \label{Fig:r_v}
   \end{figure}

After subtracting the systemic and rotational velocities we are left with the residual velocities, $v_{\rm residual}$, listed in the fourth column of Table \ref{table:derived_vel}. Given that M83 is a face-on galaxy ($i$=24$\degr$), these velocities encompass primarily the vertical velocity component. We inspect the residual velocities and find that the subtraction of these two components removes all trace of systematic behaviour from the velocities, i.e., no sinusoidal trend in the velocity residuals as a function of $\phi$. In Figure \ref{Fig:r_v} we plot the velocity residuals against  galactocentric distances for M83 including the star clusters studied in \citet{her18a} with X-Shooter. We find a dispersion in the residuals of the order of 18 km s$^{-1}$, typical of  face-on spiral galaxies \citep{her09}. Moreover, in a study focused on the central parts of M83 ($<$ 500 pc, where star formation is expected to be the highest), using star clusters, \citet{hou08} find an absence of disturbed kinematics (their Figure 8) with stellar velocities similar to our measured values.

\begin{table}
\caption{Metallicities for YMCs in M83}
\label{tab:1}
 \centering 
\begin{tabular}{cccc} 
 \hline  \hline \
Cluster & 1290--1430 \r{A}& 1435--1520 \r{A} & 1625--1790 \r{A}\\
& (dex) & (dex) & (dex) \\
  \hline\
M83-1 & $+$0.721 $\pm$ 0.315& $+$0.431 $\pm$ 0.169& $+$0.734 $\pm$ 0.214\\
M83-2 & $+$0.106 $\pm$ 0.150& $-$0.406 $\pm$ 0.116& $+$0.079 $\pm$ 0.164\\
M83-3 & $+$0.484 $\pm$ 0.076& $+$0.115 $\pm$ 0.053& $+$0.597 $\pm$ 0.071\\
M83-4 & $+$0.283 $\pm$ 0.133& $-$0.242 $\pm$ 0.108 & $+$0.166 $\pm$ 0.145\\
M83-5 & -- & -- & -- \\
M83-6 & $+$0.197 $\pm$ 0.143& $-$0.100 $\pm$ 0.064 & $+$0.277 $\pm$0.087 \\
M83-7 & $-$0.021 $\pm$ 0.259& $+$0.138 $\pm$ 0.153 & $+$0.580 $\pm$ 0.238 \\
M83-8 & $+$0.229 $\pm$ 0.056& $-$0.251 $\pm$ 0.047& $+$0.001 $\pm$ 0.067\\
M83-9 & $-$0.288 $\pm$ 0.163& $-$0.426 $\pm$ 0.076& $-$0.159 $\pm$ 0.119\\
M83-10 & $+$0.294 $\pm$ 0.062& $-$0.026 $\pm$ 0.040 & $+$0.489 $\pm$ 0.050\\
M83-11& $+$0.160 $\pm$ 0.097& $-$0.148 $\pm$ 0.062 & $+$0.058 $\pm$ 0.100\\
M83-12 & $+$0.352 $\pm$ 0.071& $-$0.088 $\pm$ 0.047& $+$0.308 $\pm$ 0.063\\
M83-13 & $-$0.200 $\pm$ 0.110 & $+$0.064 $\pm$ 0.068& $+$0.257 $\pm$ 0.102\\
M83-14 & $-$0.156 $\pm$ 0.229& $+$0.066 $\pm$ 0.105& $+$0.505 $\pm$ 0.199\\
M83-15 & $-$0.269 $\pm$ 0.327& $-$0.077 $\pm$ 0.122& $+$0.007 $\pm$ 0.199\\
M83-16 & $+$0.208 $\pm$ 0.093 & $+$0.021 $\pm$ 0.063& $+$0.495 $\pm$ 0.094\\
 \hline 
 \end{tabular}
\end{table}

%________________________________________________________________
\section{Discussion}\label{discuss}
\subsection{M83-2: Optical vs. UV}\label{sec:m83-2}
The YMC work by \citet{her18a} applying the L12 method on X-Shooter observations (optical wavelength range) and the analysis shown here utilizing the same method on COS observations (UV wavelength coverage) have only one cluster in common, M83-2. This star cluster is located at a galactocentric distance of $R/R_{25}= 0.06$.  For M83-2 \citet{her18a} measure an overall metallicity of [Z] = $+$0.17 $\pm$ 0.12 dex, compared to the metallicity measured here of [Z] = $-$0.14 $\pm$ 0.17 dex. The values extracted from these two studies covering entirely different wavelengths agree within 2$\sigma$. The difference in the measurements could be strongly influence by the uncertainties in the age estimates (see Section \ref{sec:input}). However, given that these two studies have a single target in common, limited information can be extracted from a direct comparison. A larger cluster sample with measurements in both the UV and optical is needed to draw firmer conclusions.

\subsection{Central metallicity}\label{sec:cen}
\begin{figure}
   %\resizebox{\hsize}{!}
   \centering
            {\includegraphics[scale=0.47]{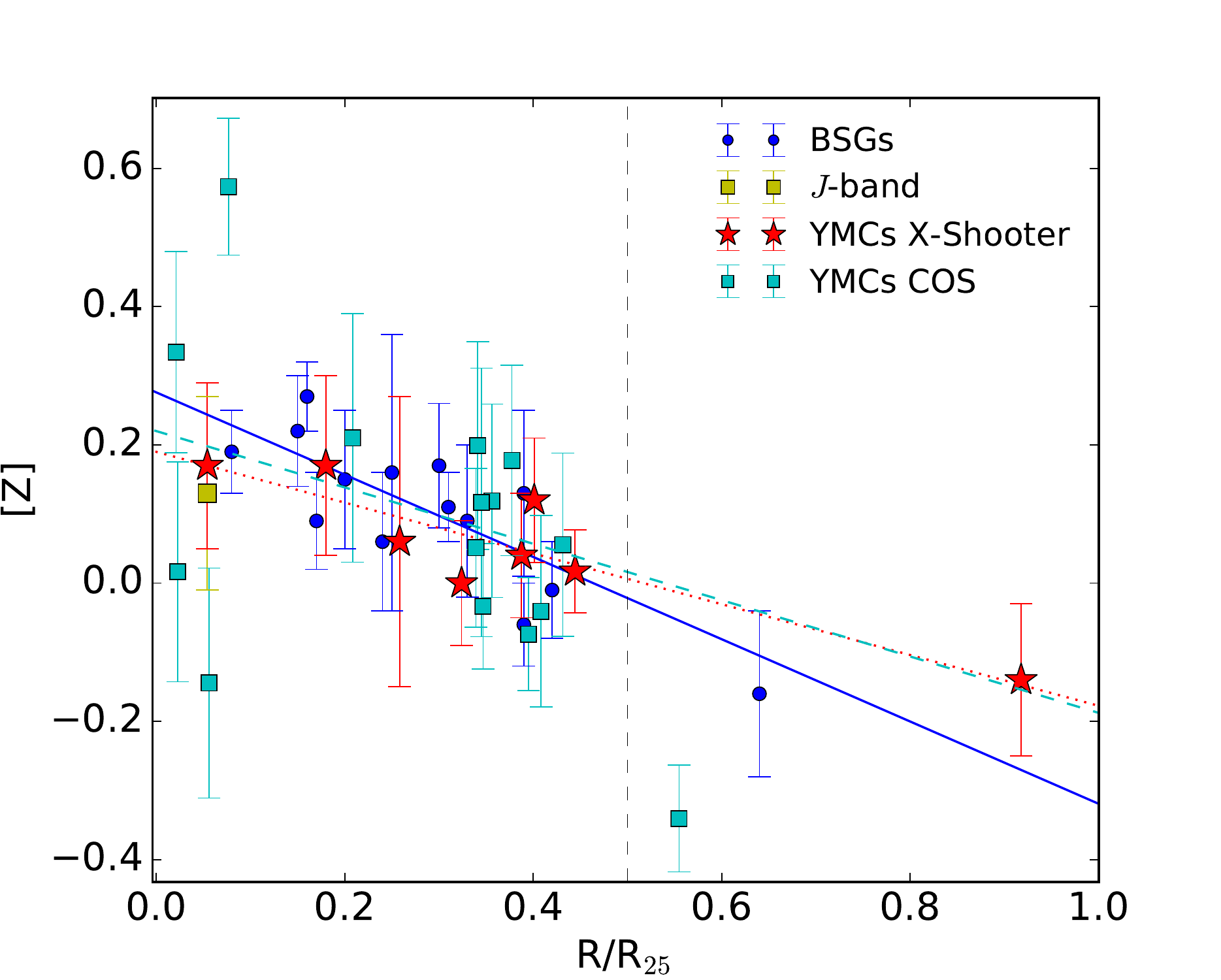}}
      \caption{Overall stellar metallicities, [Z], as a function of galactocentric distance normalised to isophotal radius, $R_{25}$. We show in cyan squares the metallicities obtained through the analysis of COS observations (this work). In yellow we display the metallicity measurement by \citet{gaz14} for cluster M83-2. In red stars we display the metallicities measured from the YMC sample observed with X-Shooter \citep{her18a}. In blue circles we include the BSG measurements \citep{bre16}. All of the measurements in this plot have been corrected to a uniform solar abundance scale as described in Section \ref{sec:cen}. Cyan dashed, red dotted, and blue solid lines show a first-order polynomial fit for $R/R_{25}< 0.5$ to the COS YMCs, X-Shooter YMCs, and BSGs, respectively. Dashed vertical line marks $R/R_{25}=0.5$.}
         \label{Fig:r_z}
   \end{figure}

In Figure \ref{Fig:r_z} we show our inferred metallicities in cyan squares as a function of galactocentric distance normalised to isophotal radius, $R_{25}$. We estimate the galactocentric distances adopting $R_{25} = 6.44$\arcmin = 9.18 kpc \citep{dev91}, $i=24$ \degr, and PA = 45 \degr \citep{com81}. For comparison to our UV IL metallicities we include in Figure \ref{Fig:r_z} the BSG metallicities by \citet{bre16}, the YMC measurement by \citet{gaz14}, and the IL optical metallicities by \citet{her18a}. \par
We note that to compare the different studies, similar to what was done by \citet{her18a}, we homogenise the sets of measurements to a single abundance scale. In our analysis we adopt a Solar composition as published by  \citet{gre98} with a metallicity mass fraction of Z$_{\rm YMC}$ = 0.0169. The analysis of the BSGs by \citeauthor{bre16} adopts a Solar oxygen abundance by \citet{asp09}, and Solar composition for the rest of the elements by \citet{gre98} with a total metallicity mass fraction of Z$_{\rm BSG}$ = 0.0149. We revise the BSG metallicities following
\begin{equation}
[Z]_{\rm YMC} = [Z]_{\rm BSG} - \log\bigg(\frac{Z_{\rm YMC}}{Z_{\rm BSG}}\bigg) = [Z]_{\rm BSG} - 0.06
\end{equation}
\citet{gaz14} measured the metallicity of a YMC in M83 using the $J$-band method. Given that in stellar clusters with ages $\simeq$7 Myr the most massive stars in the cluster, RSGs, dominate the integrated light in the near-IR regime one can analyse their spectra by approximating the IL to a single RSG. The $J$-band analysis uses \textsc{MARCS} models based on Solar abundances from \citet{gre07} with a total metallicity mass fraction of Z$_{\rm Jband}$ = 0.012. To directly compare our metallicities to that inferred by \citeauthor{gaz14} we adopt the following relation:
\begin{equation}
[Z]_{\rm YMC} = [Z]_{\rm Jband} - \log\bigg(\frac{Z_{\rm YMC}}{Z_{\rm Jband}}\bigg) = [Z]_{\rm Jband} - 0.15
\end{equation}

We note that going forward the metallicities from the studies discussed here are on the same abundance scale. In addition to the work of \citet{her18a}, the metallicity of M83-2 has also been obtained by \citet{gaz14} using the $J$-band method. \citeauthor{gaz14} measure an overall metallicity of [Z]=$+$0.13 $\pm$ 0.14 dex, a value within the errors of that obtained by \citet{her18a} with X-Shooter. We find that the metallicity inferred here for M83-2, [Z] = $-$0.14 $\pm$ 0.17, is consistent within the uncertanties from the $J$-band metallicity. \par
We now compare the overall metallicities obtained strictly within the central regions of the galaxy. We arbitrarily chose galactocentric distances of $R/R_{25}<$0.1 given that all the different studies (BSG, $J$-band, X-Shooter, and COS) have a minimum of one metallicity measurement within this range. In our YMC sample we find 4 targets with radii $R/R_{25}<$0.1, M83-1, M83-2, M83-3 and M83-4. The average metallicity of these clusters is [Z] = $+$0.20 $\pm$ 0.15 dex. This measurement is within the errors of the BSG at a galactocentric distance of $R/R_{25}\sim$0.08 studied by \citet{bre16} with a metallicity of [Z] = $+$0.25 $\pm$ 0.06 dex. A similar agreement is observed with the YMCs by \citeauthor{gaz14} and the measurement by \citeauthor{her18a} discussed above. We note, however, that although the averaged abundances derived from the inner most COS YMCs agree well with other studies, we observe a large scatter, $\sigma_{\rm STD}$=0.27 dex, in the metallicity measurements of these four clusters. With a metallicity of [Z] = $+$0.57 $\pm$ 0.10 dex, YMC M83-1 is the most discrepant target, when compared to the rest of the clusters. The lower uncertainties, $-\sigma_{\rm Age}$, introduced by the age estimates for this specific cluster may vary its derived metallicity by $\lesssim$0.4 dex, whereas the upper uncertainties, $+\sigma_{\rm Age}$, can change the derived metallicity by $<$0.1 dex (see Table \ref{table:input}). We believe that one of the main components possibly driving the observed scatter in our UV metallicities are the age estimates. This behavior in the upper uncertainties of the UV metallicities is similar to the optical values, where \citet{her18a} estimated the overall metallicities of their YMC sample may vary on average by $\sim$0.1 dex, depending on the age uncertainties. Improved age measurements could help reduce the dispersion in our UV metallicity measurements. \par

\subsection{The Inner Disc Metallicity Gradient}\label{sec:gradient}
Overall, Figure \ref{Fig:r_z} shows that the best agreement in the measurements from the different approaches, IL (optical and UV), BSG, and $J$-band is found at $R/R_{25}\sim$0.1-0.5, and significant scatter at  $R/R_{25}<$0.1. In this Section we examine the metallicity gradient observed in the different studies for objects at galactocentric distances $R/R_{25}< 0.5$. To quantify the metallicity gradient in Figure \ref{Fig:r_z} we show linear regressions to the BSGs (solid blue line), YMCs observed with X-Shooter (red dotted line), and the YMCs studied with COS here (cyan dashed line). The resulting linear fits, weighted by the uncertainties in the measurements, are as follows:
\begin{equation}
[Z]_\mathrm{BSG} = -0.60\: (\pm 0.19) \: R/R_{25} + 0.20\: (\pm 0.05)
\end{equation}
\begin{equation}
[Z]_\mathrm{X-Shooter} = -0.37\: (\pm 0.29) \:R/R_{25} + 0.19\: (\pm 0.09)
\end{equation}
\noindent and
\begin{equation}\label{eq:cos_grad}
[Z]_\mathrm{COS} = -0.38\: (\pm0.20)\: R/R_{25}\: + 0.20\: (\pm0.06)
\end{equation}
The best fit slopes agree within the errors in all of the studies, displaying a clear inner disc gradient previously observed also in \ion{H}{2} region measurements \citep{bre02, bre05, bre09}. Additionally, we note that a comparison between the slopes for the X-Shooter measurements and those from this work, COS measurements, are generally consistent, closely overlapping each other as shown in Figure \ref{Fig:r_z}. This shows that the IL analysis of such young populations provides consistent results irrespective of the wavelength range. Lastly, the BSG gradient of $-$0.065 $\pm$ 0.021 dex kpc$^{-1}$ is slightly steeper than those measured in the X-Shooter and COS observations, $-$0.040 $\pm$ 0.032 dex kpc$^{-1}$ and $-$0.041 $\pm$ 0.022 dex kpc$^{-1}$ respectively. %Combining the nebular abundances measured from the [\ion{N}{ii}/\ion{O}{ii}] ratio by \citet{bre07} with their own measurements of five \ion{H}{ii} regions \citet{bre09} measured an inner disc ($R/R_{25}< 1.4$) metallicity gradient of $-$0.030 $\pm$ 0.002 dex kpc$^{-1}$, a slightly shallower gradient than that obtained here. The difference in gradient estimates most likely originates from the fact that the galactocentric coverage in the measurements by \citet[][$R/R_{25}< 1.4$]{bre09} is slightly different than those used in the X-Shooter and COS measurements ($R/R_{25}< 0.5$). We also note that these nebular abundances from the [\ion{N}{ii}/\ion{O}{ii}] ratio have an offset of $\sim$0.1 dex towards lower metallicities with respect to the stellar metallicities from both, YMCs and BSGs. \par

For galactocentric distances $>0.5\:R_{25}$ we are limited to only three data points. From Figure \ref{Fig:r_z} is unclear if the abundance gradient remains steep beyond $0.5\:R_{25}$, or if it changes to a relatively flat one. Given the low number of stellar measurements beyond $0.5\:R_{25}$ it becomes a challenge to draw any concrete conclusions on the metallicity trends in the outer parts of the galaxy. Additional stellar abundances in this galactocentric regime will help discriminate between a single or multiple gradient fit. We present a more detailed discussion of the gradient at distances of $>0.5\:R_{25}$ in Section \ref{sec:ste_neb} combining stellar and nebular measurements.\par
%The agreement between the COS metallicity gradient and that from independent studies, including the X-Shooter measurements also confirm the reliability of the L12 method when applied to UV observations of young populations. We note, however, that the accuracy of the L12 technique amply relies on the input accuracy, i.e. isochrones and line lists. In a detailed abundance study of the IL of the GC G280 in M31, \citet{lar18b} found significant differences when comparing abundances obtained from optical and infrared  observations using the line lists by \citet{cas04}. They find differences between the optical and infrared values of Fe abundance of 0.34 dex. \citet{lar18b} reduce these differences by changing their line lists to the most recent version of the Kurucz atomic line lists\footnote{http://kurucz.harvard.edu/linelists.html}. Replacing the \citet{cas04} lists produces the most significant changes in their infrared measurements, in contrast to those observed in the optical abundances. \citet{lar18b} emphasize that the optical abundances are in generally good agreement, it is the infrared abundances for which the different line lists show significant changes. Looking at Figure ~\ref{Fig:r_z} we can see that the differences of the order of $\sim$0.3 dex observed in the work of \citet{lar18b} between abundances in the optical and infrared are clearly absent in our optical/X-Shooter and the UV/COS studies.  \par

\subsection{Mass-Metallicity Relation (MZR)}\label{sec:MZR}
As described in Section \ref{sec:intro}, the MZR has become an essential tool for understanding the formation and evolution of starburst galaxies. For many decades, since the original observations by \citet{leq79}, the MZR has been scrutinised primarily through the analysis of \ion{H}{2} regions. Well known complications arise when comparing the MZRs obtained through different methods (T$_{e}$-based method or strong-line method), where different techniques yield metallicities with differences as high as $\sim$0.7 dex \citep{bre04, gar04}. This means that the quantitative shape of the MZR varies depending on the method used in obtaining such measurements. The exact cause of these differences is not entirely understood, although one possibility is systematic errors originating in the strong-line method \citep{kew08}. \par
\begin{figure}
   %\resizebox{\hsize}{!}
            {\includegraphics[scale=0.41]{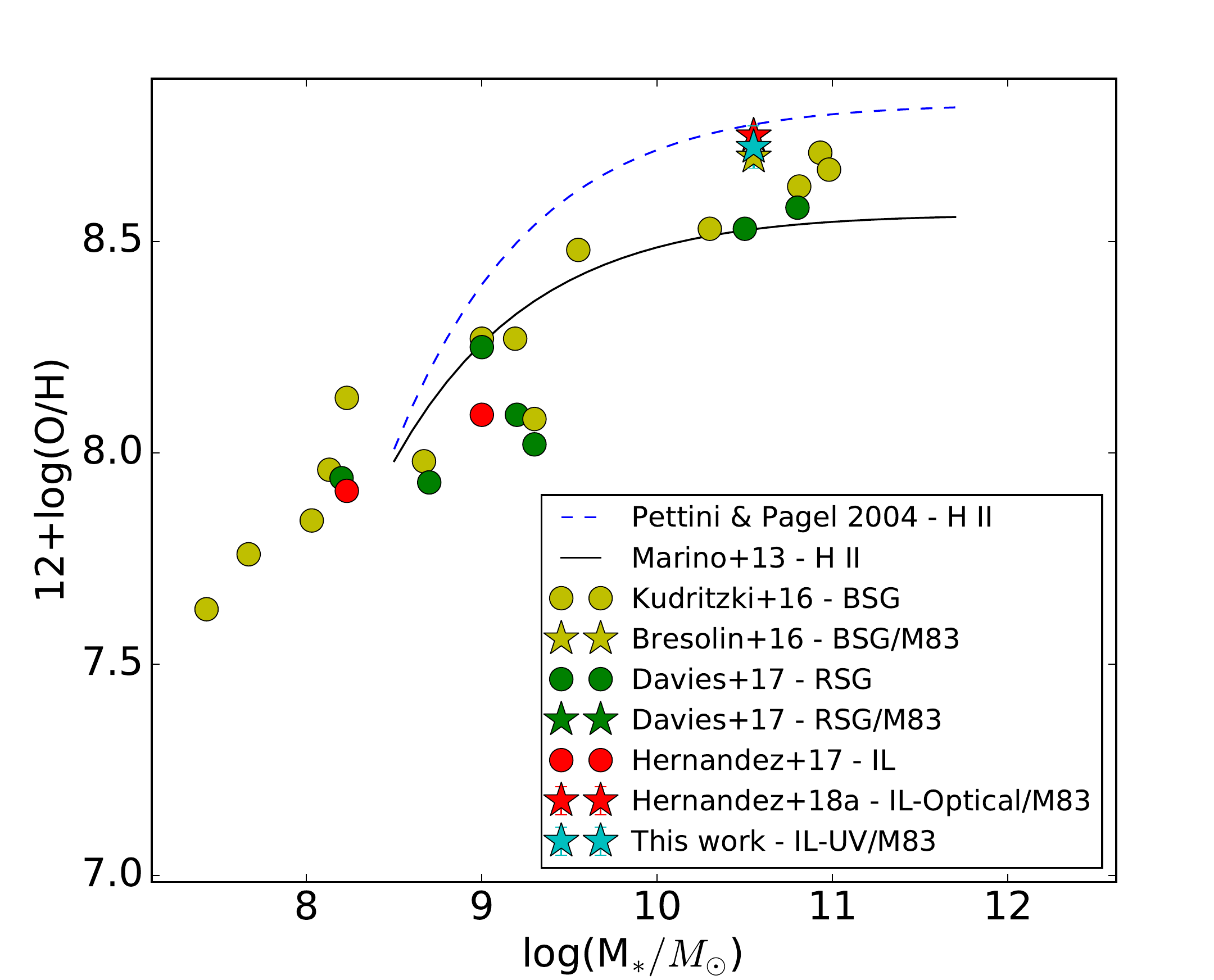}}
      \caption{Mass-Metallicity Relation as measured from BSG \citep{kud16, bre16}, RSGs \citep{dav17} and IL studies of YMCs \citep{her17, her18a} shown in yellow, green and red symbols. Shown as a cyan star we display the abundance measured in this work from the UV observations. We also include in dash and solid lines the relations obtained by \citet{bar17} using the semi-empirical calibrator by \citet{pet04} and the T$_{e}$-based method by \citet{mar13} for more than 1700 MaNGA galaxies, respectively.}
         \label{Fig:mzr}
   \end{figure}

In Figure \ref{Fig:mzr} we display in dashed and solid lines the MZRs obtained by \citet{bar17} using the strong-line (semi-empirical) calibrator by \citet{pet04} and the T$_{e}$-based method by \citet{mar13}, respectively. The study by \citet{bar17} focused on more than 1700 galaxies in the Sloan Digital Sky Survey (SDSS)-IV Mapping Nearby Galaxies at APO \citep[MaNGA;][]{bun15} determining metallicities at the same physical scale (effective radius, R$_{\rm eff}$). Comparing these two MZRs one can clearly see an offset between the two relations particularly towards higher $\log$(M$_*$/M$_\odot$). We point out that the work of \citet{bar17} covered $\log$(M$_*$/M$_\odot$) values between 9 and 11.5. In addition to these nebular MZRs, we also show in Figure \ref{Fig:mzr} as yellow, green and red circles the stellar metallicities obtained through the analysis of BSG (compilation by \citealt{kud16}), RSG (compilation by \citealt{dav17}) and IL of YMCs \citep[][mass estimates from \citealt{ryd95} and \citealt{rom06}]{her17}, respectively. Although low in number, the IL metallicities are well within the scatter of the BSG and RSG metallicities. Similarly, one can see that  the MZR obtained from the T$_{e}$-based method \citep{mar13} is in slightly better agreement with the stellar MZR than the one inferred from the strong-line/semi-empirical method \citep{bre16, kud16, dav17, her18a}. Extrapolating from the MZRs by \citet{bar17} for those galaxies with $\log$(M$_*$/M$_\odot$)$<$8.5 we clearly see a deviation from the MZRs for both methods, however, this is most likely caused by the mass range covered in the work of \citet{bar17}. \par
Assuming a stellar mass of $\log(\rm M_{*}/\rm M_{\odot})$= 10.55 \citep{bre16} we include in Figure \ref{Fig:mzr}, as star symbols, the stellar metallicities inferred through independent methods for M83: BSG \citep[yellow,][]{bre16}, RSG \citep[green,][]{dav17}, IL-Optical \citep[red,][]{her18a} and IL-UV (cyan, this work). We note that these studies measure overall metallicities, [Z], of either individual stars or a population of stars. We then convert these [Z] values to oxygen abundances adopting the solar oxygen composition by \citet{asp09}, 12+log(O/H) = 8.69. Additionally, for spiral galaxies which generally show metallicity gradients it is common to adopt a characteristic metallicity typically measured at $R= 0.4 \:R_{25}$ which is a good surrogate for the integrated metallicity of the whole galaxy \citep[][this is the metallicity obtained from spatially unresolved spectroscopy observations of a galaxy]{zar94, mou06}. We take the average metallicities of YMCs M83-8, M83-13, M83-15 and M84-16, all of which are located at $R\sim 0.4\: R_{25}$ and obtained an average oxygen abundance of 12+log(O/H)= 8.72 $\pm$ 0.05 dex (shown in Figure \ref{Fig:mzr}). A similar abundance is found when using eq. ~\ref{eq:cos_grad} to estimate the metallicity of M83 at $R= 0.4 \:R_{25}$.\par
The main objective of Figure \ref{Fig:mzr} is to show the agreement between the metallicities obtained through the different stellar methods. More significantly, in the four measurements of M83 (star symbols) we can see that these overlap with each other, validating the methods used in the the stellar MZR. To quantify the scatter in the stellar metallicities we fit a linear regression to all of the points in Figure \ref{Fig:mzr}. We find a standard deviation of the residuals (after subtracting the best linear fit to the data) of 0.104 dex. The scatter in the stellar measurements is slightly larger than those estimated by \citet{bar17} of 0.097 and 0.067 dex for the \citet{pet04} and \citet{mar13} calibrators, respectively. We note that the stellar relation is based on a sample size orders of magnitude smaller than those by \citet{bar17}. However, Figure \ref{Fig:mzr} provides independent insight into the MZR based on abundance measurements from young populations (from both YMCs and BSGs) unaffected by the systematic uncertainties accompanying the nebular abundances. \par

  \begin{figure*}
   %\resizebox{\hsize}{!}
            {\includegraphics[scale=0.45]{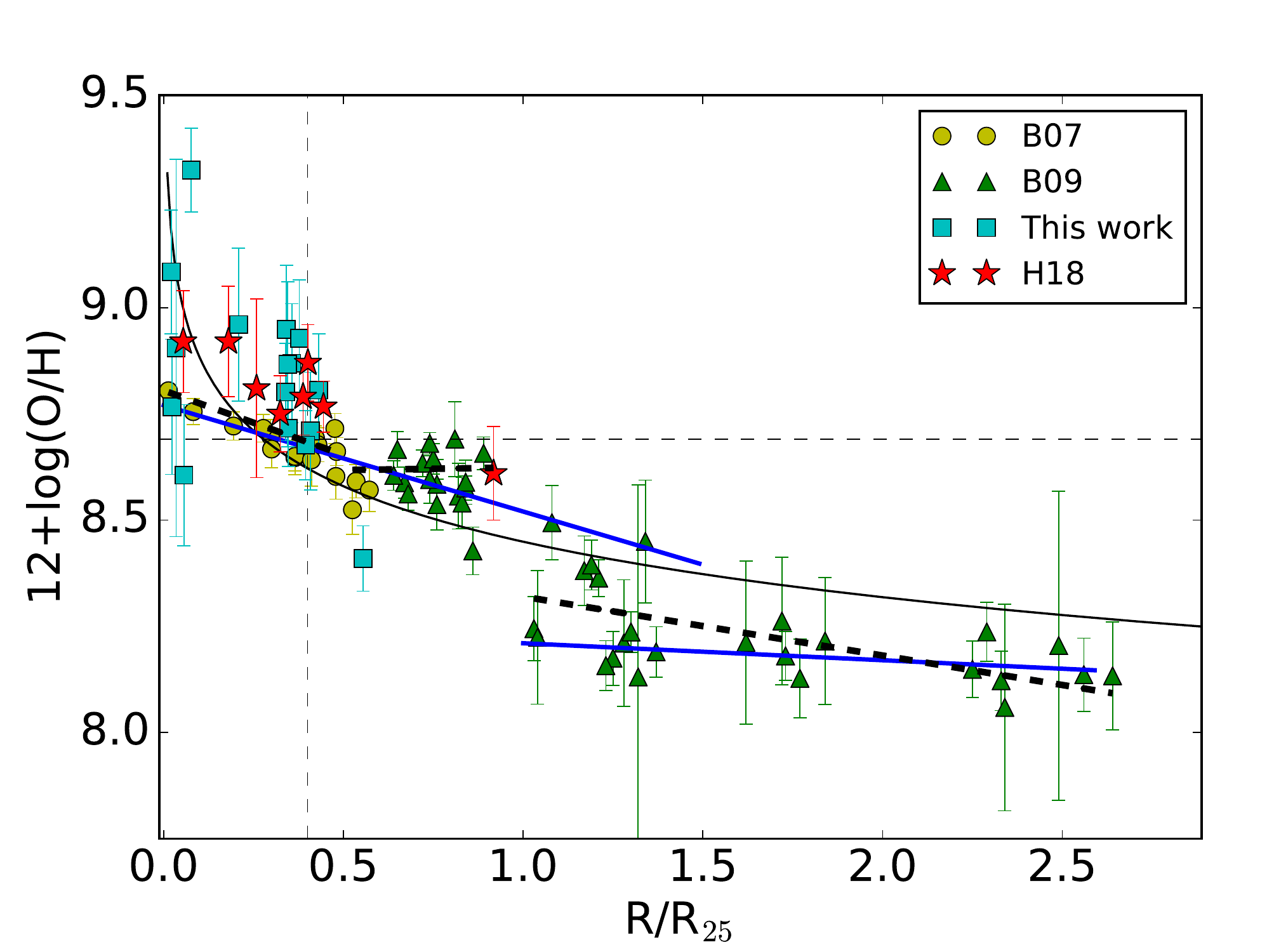}}
             {\includegraphics[scale=0.45]{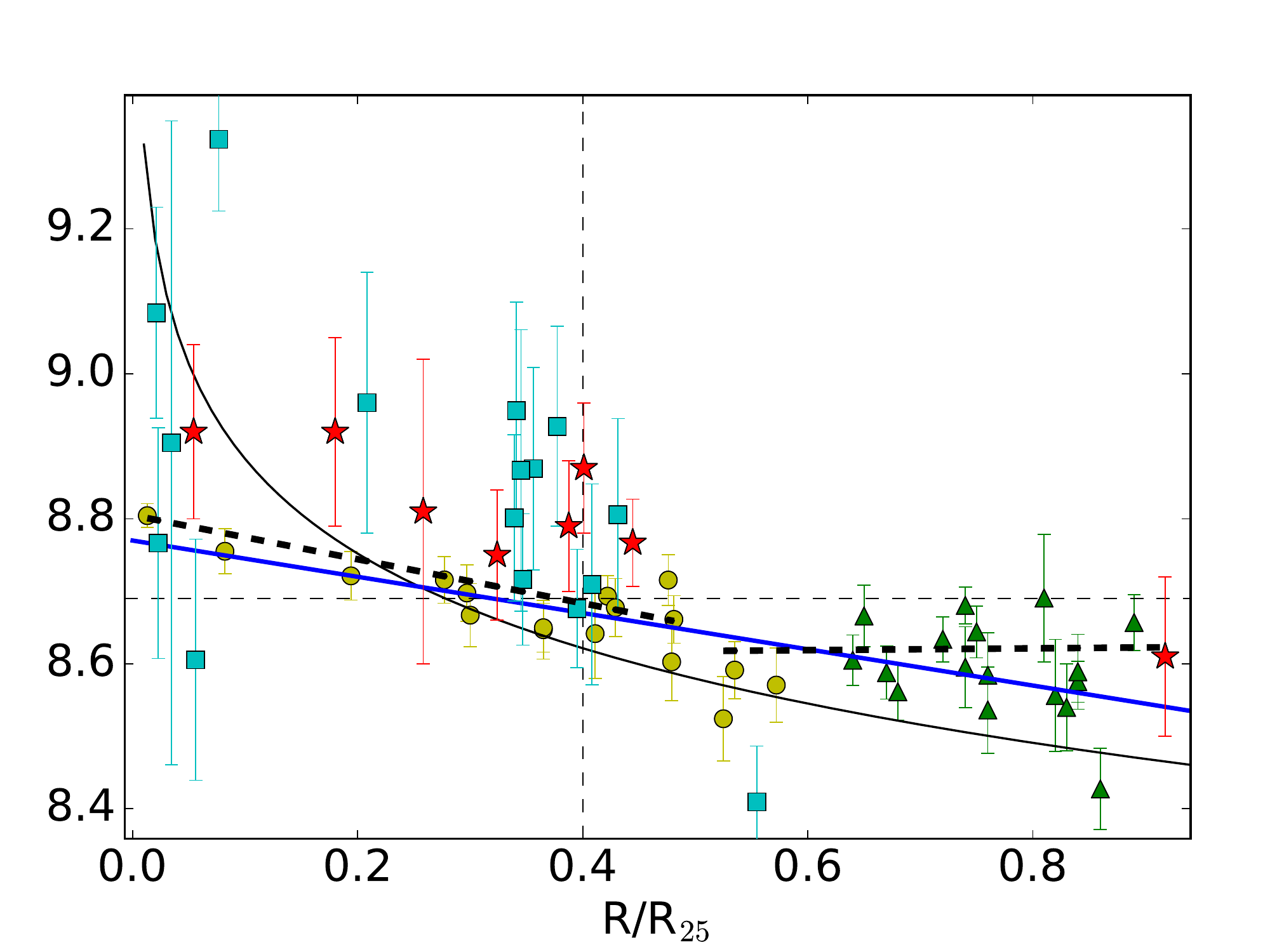}}
      \caption{Left panel: Oxygen abundance as a function of galactocentric distance normalised to isophotal radius, $R_{25}$. Right panel: Zoomed-in version of left panel. In cyan squares we display the overall metallicities measured for the YMC studied in this work, converted to oxygen abundances. In red stars we show the YMC measurements by \citet{her18a}. In yellow circles we present the \ion{H}{2} abundances from \citet{bre07}. In green triangles we show the \ion{H}{2} measurements from \citet{bre09}. In thick dashed black lines we show linear regressions at three different galactocentric distance ranges: $<0.5\:R_{25}$, 0.5-1.0$\:R_{25}$, and $>1.0\:R_{25}$. For comparison we show a single power-law fit as a thin black line and the linear regressions estimated by \citet{bre09} as solid blue lines.} We mark with  thin dashed horizontal and vertical lines the adopted solar oxygen abundance and the galactocentric distance $R = 0.4\: R_{25}$. 
         \label{Fig:r_oh}
   \end{figure*}

\subsection{Stellar vs. Nebular}\label{sec:ste_neb}
We now compare our stellar abundances to previously published abundance determinations using nebular lines. We point out that these abundances use the same Solar composition as that from the work of \citet[][hereafter B07]{bre07} and \citet[][hereafter B09]{bre09}. The main driver for comparing our stellar measurements to those from B07 and B09 is the sample size. The combination of these studies provide a total of 59 data points. These measurements were obtained using [\ion{N}{2}/\ion{O}{2}] ratios along with the empirical calibration by \citet{bre07} based on \ion{H}{2} regions with available T$_{e}$-based abundances.\par

In \citet{her18a}, we compare the metallicities from the YMCs observed with X-Shooter to several studies adopting different nebular calibrations based on both theoretical and empirical data. We find that the best agreement between the stellar metallicities and the nebular abundances is obtained using the strong line metallicity indicator [\ion{N}{2}/\ion{O}{2}] ratio and the calibration by B07, in spite of the consistent offset towards lower metallicities from the stellar ones. In Figure \ref{Fig:r_oh} we plot the oxygen abundances as a function of galactocentric distance. Comparing the stellar measurements obtained in this work (cyan squares in Figure \ref{Fig:r_oh}) we find a similar behaviour, where the stellar abundances are slightly higher than those inferred from the nebular lines using the B07 calibration (yellow circles in Figure \ref{Fig:r_oh}). %The best agreement between the stellar and nebular abundances is observed at $R= 0.4\: R_{25}$.
\par
In addition to comparing our stellar abundances, both from X-Shooter and COS, to the abundances by B07 we also include in Figure \ref{Fig:r_oh} the \ion{H}{2} region measurements from B09 as green triangles. In their work B09 focus their study on the extreme outer disc of M83 ($R>R_{25}$). They report on two different radial gradients, where the first one extends to the optical edge of the disc, $R\sim R_{25}$, and the second one covers $R>R_{25}$. B09 find that the outer abundance measurements flatten to an approximately constant value, or at least a less steep gradient. \par

\subsection{Gradient Breaks}
As mentioned in Section \ref{sec:gradient}, the limited number of stellar metallicities at $>0.5\:R_{25}$ prevented us from identifying any changes in the metallicity gradient. By combining our stellar metallicities with those from nebular studies we increase the number of data points at galactocentric distances $>0.5\:R_{25}$. From Figure \ref{Fig:r_oh} (right panel) we can observe a possible change in the abundance trend in the B09 measurements for $R\sim 0.5$-$1.0\:R_{25}$. To quantify these gradients, we divide the abundance measurements in three bins: $<0.5\:R_{25}$, 0.5-1.0$\:R_{25}$, and $>1.0\:R_{25}$. The bins were arbitrarily chosen based on the corotation radius ($\sim$0.44 $\:R_{25}$), and the edge of the optical disc ($\sim$ 1 $\:R_{25}$). We apply a linear regression to the different bins which include both the stellar and nebular metallicities, and display the fits in Figure \ref{Fig:r_oh} as dashed thick black lines. We find a gradient of $-0.303\pm0.017$ dex $R_{25}$$^{-1}$ for the central parts of the disc, $0.012\pm0.088$ dex $R_{25}$$^{-1}$ for the region close to the optical edge of the disc, and $-0.139\pm0.037$ dex $R_{25}$$^{-1}$ for the outer regions of the disc. These slopes correspond to $-0.033\pm0.002$ dex kpc$^{-1}$, 0.001$\pm$0.010 dex kpc$^{-1}$ and $-0.015\pm0.004$ dex kpc$^{-1}$.  Clearly the steepest gradient is found in the inner regions of the M83 disc, this value is comparable to that obtained in the linear regression applied to the stellar metallicities alone in Section \ref{sec:gradient}, $-0.041 \pm 0.022$ dex kpc$^{-1}$. Close to the optical edge, $R= 0.5$-$1.0\: R_{25}$, we find a relatively flat gradient, which beyond the optical edge disc goes back to a slightly steeper slope again. As a general comparison we show in Figure \ref{Fig:r_oh} a single power-law fit as a thin solid line. We find that the standard deviation of the residuals for the single power-law fit is higher (0.27 dex) than that from the multiple linear regression (0.12 dex). We contrast these values with the standard deviation of the residuals for the linear regressions as calculated by \citet[][their equations 1 and 2]{bre09} which we estimate to be 0.26 dex. The residuals are estimated by subtracting the best fit to the empirical data. The lower dispersion in the residuals from the multiple linear regression presented here (thick dashed lines in Figure \ref{Fig:r_oh}) seems to suggest that a three-component fit best describes the gradient trends as observed from the full set (stellar+\ion{H}{2}).  \par

If confirmed to be genuine, the inner break observed in the stellar and nebular abundances of M83 is similar to that reported by \citet{roy97}, who found a break in the abundance gradient of the barred spiral galaxy NGC 1365 around $R\sim0.55 \:R_{25}$. For comparison, the bar semi major axis in NGC 1365 has a radius of $\sim$ 0.36 $R_{25}$. According to their study, beyond $R\sim0.55 \:R_{25}$ the abundance trend becomes flat. Given the presence of the break in the radial abundance distribution, \citet{roy97} propose that the bar may have originated in recent times ($<$1 Gyr) due to an interaction or a merger.
  \citet{fri94} and \citet{fri95} explain the development of breaks in the radial abundance distribution. In their three-dimensional numerical work \citeauthor{fri94} and \citeauthor{fri95} state that a break in the metallicity gradient of a galaxy can occur changing from \textit{steep} to \textit{shallow}. The former is due to vigorous enrichment by star formation episodes taking place in the bar, the latter caused by dilution effects (i.e., mixing low-metallicity material with enriched gas) of the outward flow of interstellar gas induced by the non-axisymmetric potential bars \citep{sel93} beyond the corotation radius. This is the case for NGC 3359, a barred spiral galaxy displaying a break in the abundance gradient, going from steep in the inner regions of the galaxy to flat beyond $R\sim 0.5 \:R_{25}$. Based on the study of \citet{mar95}, a quick phase of intense star formation took place along the bar major axis of NGC 3359 and its spiral arms; this galaxy is assumed to have a relatively young bar ($<1$ Gyr). \par
In one of the largest and most homogenous studies of \ion{H}{2} regions performed to date, \citet{san14} analyse observations of more than 7000 regions observed by the Calar Alto Legacy Integral Field Area \citep[CALIFA, ][]{sanc12} survey. \citet{san14} find that galaxies \textit{without} clear evidence of an interaction display a common gradient in their oxygen measurements with a characteristic slope of $-$0.1 dex $r_{e}^{-1}$ between 0.3 and 2 disc effective radii ($r_{e}$). According to this work this single slope is independent of mass, bar, morphology or absolute magnitude. However, those galaxies with evidence of interactions/mergers show a significantly shallower gradient. \par

According to \citet{lun04} corotation in M83 is at 2.83\arcmin or $R\sim0.44\:R_{25}$, relatively close to the proposed break in the abundance gradient at $R\sim0.5\:R_{25}$ (see right panel of Figure \ref{Fig:r_oh}). Furthermore, the abundance trends observed in NGC 1365 and in NGC 3359 are similar to what we find in M83, including the location of the breaks which are found at similar galactocentric distances possibly implying: (i) the bar in M83 may have formed in the last Gyr as a result of an interaction or merger, (ii) the inner regions of the galaxy have experienced vigorous star formation injecting enriched material into the central parts of M83, (iii) a flow from the centre of the galaxy outward, beyond the corotation radius, is taking place diluting the less enriched matter with high metallicity material at $R>0.5 \:R_{25}$. \par

It is worth noting that in the Milky Way, several studies have found bimodal chemical abundance distributions. A clear discontinuity in the radial distribution of abundances of Cepheids has been observed at galactocentric distances $>$ 10 kpc \citep{and04, yon06}. Theoretical simulations show that if the spiral structure is quasi-stationary, it is the location of the corotation radius that dictates the final abundance radial distribution \citep{ach05}. In the case of the Milky Way, the lack of efficient radial mixing at galactocentric distances beyond the corotation radius could be the factor responsible for the discontinuity in the abundance gradient observed in the Cepheid studies.

B09 find an abrupt discontinuity at the edge of the optical disc. Unfortunately, given that all of our targets, both in the optical and UV study, lie at galactocentric radii $R<1.0\: R_{25}$ we cannot confirm this discontinuity with our IL study. However, we now discuss the outer disc abundance gradient as observed from the measurements by B09 (see left panel of Figure \ref{Fig:r_oh}). Models by \citet{fri94} and \citet{fri95} predict an outer break in the radial abundance distribution of galaxies going from \textit{shallow} to \textit{steep} which defines how far the outward flow has penetrated the outer disc. These models show that this second break moves to larger radii with time. As stated above, taking all the abundance measurements by B09 with $R>1.0 \:R_{25}$ we measure an abundance gradient of $-$0.015 $\pm$ 0.004 dex kpc$^{-1}$, this preceded by a flatter gradient of 0.001 $\pm$ 0.010 dex kpc$^{-1}$. We note that B09 instead find a flatter outer-disc gradient of $-$0.005 dex kpc$^{-1}$, the difference arises from the fact that B09 excludes the five most metal rich measurements at $R>1.0 \:R_{25}$ when applying the liner regression to the outer disc abundances as they assume these belong to the group of inner \ion{H}{2} regions. The abundance gradient pattern observed here we believe displays a  \textit{shallow}-to-\textit{steep} gradient trend, similar to what is predicted by \citeauthor{fri94} Assuming this second break in the abundance gradient is genuine, we can then propose that the radial flow has penetrated the outer disc as far as $R\sim1.0 \:R_{25}$.\par

%__________________________________________________

\section{Conclusions}\label{con}
To complement the invaluable contributions made from the analysis of \ion{H}{2} regions when investigating the present-day chemical state of starburst galaxies, we seek alternative tools to avoid the characteristic problems arising when comparing metallicities obtained through different nebular methods \citep{bre08, kew08}. Star clusters provide a viable alternative to the analysis of \ion{H}{2} regions. 
In this work we extend our initial M83 metallicity study \citep{her18a} adding 15 more YMCs to our original sample. As stated in \citet{her18a}, this type of analysis is especially critical for environments with metallicities above solar since some nebular methods fail to provide accurate measurements by underestimating the gas-phase abundances \citep{sta05, sim11, zur12}. \par

In this paper we present the first IL study of YMCs focused on the UV component of 15 targets observed with the COS spectrograph. This in contrast to the work done in \citet{her18a}, where we analyse the IL of 8 YMCs observed in the optical wavelength regime with the X-Shooter spectrograph. We found a single YMC (M83-2) in common between the optical and UV work. Comparing the overall metallicities obtained from the two independent studies we find that the COS measurement is well within 2$\sigma$ from the X-Shooter metallicity. Additionally, averaging the metallicities of four different YMCs in the central regions of M83 ($R<0.1\:R_{25}$) we estimate [Z] = $+$0.20 $\pm$ 0.15 dex, a measurement in agreement with those obtained through independent methods by \citet{gaz14}, \citet{bre16}, and \citet{her18a} for YMCs and BSGs at a galactocentric distance of $R\sim0.1 \:R_{25}$. This agreement supports the above solar metallicity in the central regions of M83 as well as the reliability of the L12 method when applied to young populations observed in both the optical and the UV wavelength regimes. We also compare the metallicity gradient obtained from our initial X-Shooter YMC sample with that from the COS YMC observations finding excellent agreement (within their uncertainties) between our two studies. \par
We combine our IL stellar abundances with those obtained from \ion{H}{2} regions by \citet{bre07} and \citet{bre09} and find two possible breaks in the abundance gradient of M83 at $R\sim0.5\:R_{25}$ and $R\sim1.0\:R_{25}$. If the abundance breaks are indeed intrinsic to M83 we raise the possibility that the metallicity gradient in this spiral galaxy follows a \textit{steep}-\textit{shallow}-\textit{steep} trend. Such a scenario is predicted by the models of \citet{fri94} and \citet{fri95} where the first gradient break is usually located near the corotation radius and is caused by recent star formation episodes and a relatively young bar (formed $<$1 Gyr through either an interaction or merger) and the flatter gradient is dominated by dilution effects due to outflows (i.e. mixing of low-metallicity material with enriched matter). The second break, according to \citeauthor{fri94}, marks the farthest galactocentric distance the outward flow has been able to penetrate, which in the case of M83, we propose is $R\sim1.0\:R_{25}$.\par
The work presented here, based on the analysis of IL observations of YMCs, demonstrates that this method can be used as an alternative tool to \ion{H}{2}-techniques, applicable not only in the optical wavelength regime, but also in the UV. By successfully expanding the stellar metallicity analysis to UV wavelengths we have made it possible to simultaneously study the interstellar medium (ISM; metallicities in the neutral ISM are typically assessed through the analysis of Far UV absorption lines arising from heavy elements using bright UV sources as background) kinematics and stellar properties. Such analysis can bridge our understanding of galactic outflows, stellar evolution and stellar feedback not only locally, but also at high redshifts ($z\gtrsim$2) where the rest-frame UV light is shifted to optical/IR wavelengths. \par

\acknowledgments

These data are associated with the HST GO programs 14681 (PI: A. Aloisi). Support for this program was provided by NASA through grants from the Space Telescope Science Institute. Some of the data presented in this paper were obtained from the Mikulski Archive at the Space Telescope Science Institute (MAST).


\vspace{5mm}
\facilities{HST(COS)}

%\software{astropy \citep{2013A&A...558A..33A}}



\appendix



\bibliographystyle{aasjournal}
\bibliography{m83} 


\end{document}